\documentclass[10pt,twocolumn]{IEEEtran}

\usepackage{amsmath,amsfonts,graphics,amssymb,epsfig,subfigure,color,cite}
\usepackage{amsthm,textcomp}
\theoremstyle{plain}
\newtheoremstyle{mystyle}
  {0mm}
  {0mm}
  {}
  {4mm}
  {\bfseries}
  {:}
  { }
  {\thmname{#1}\thmnumber{ #2}\thmnote{ (#3)}}
\theoremstyle{mystyle}
\usepackage{bm}

\usepackage{array}
\usepackage{multirow}

\usepackage{enumerate}
\usepackage{algorithm}
\usepackage{algpseudocode}
\usepackage{algcompatible}
\algblockdefx{WHILE}{ENDWHILE}[1]%
  {\textbf{while }#1 \textbf{do}}%
  {\textbf{end}}
\algblockdefx{FORP}{ENDFORP}[1]%
  {\textbf{for }#1 \textbf{do in parallel}}%
  {\textbf{end}}
\algblockdefx{noEnd}{noEndEnd}[1]
    {#1}
    {{\Statex}}
\algblockdefx{FOR}{ENDFOR}[1]{\textbf{for }#1 \textbf{do}}{}\algnotext{ENDFOR}
\algblockdefx{BULLET}{EndBULLET}{$\bullet$ }{}\algnotext{EndBULLET}
\algblockdefx{PROC}{ENDPROC}[1]
    {\textbf{procedure }#1 {}}
    {}\algnotext{ENDPROC}
\algblockdefx{IF}{ENDIF}[1]%
  {\textbf{if }#1 \textbf{then}}%
  {}\algnotext{ENDIF}
\algnewcommand\algorithmicprocedure{\textbf{procedure}}
\algnewcommand\FUNC{\item[\algorithmicprocedure]}%
\algnewcommand\algorithmicendprocedure{\textbf{end procedure}}
\algnewcommand\ENDFUNC{\item[\algorithmicendprocedure]}%
 \usepackage{setspace}
\let\Algorithm\algorithm
\renewcommand\algorithm[1][]{\Algorithm[#1]\setstretch{1.4}}
\usepackage{float}
\usepackage{makecell}
\usepackage{graphicx}
\usepackage{xcolor}
\newtheorem{thm}{Theorem}

\usepackage{bbm}

\usepackage[shortlabels]{enumitem}

\newcommand{\argmin}{\operatornamewithlimits{argmin}}

\makeatletter
\newcommand{\vast}{\bBigg@{4.5}}
\newcommand{\Vast}{\bBigg@{7.5}}
\makeatother

\allowdisplaybreaks

\begin{document}
    \title{Communication-Efficient Split Learning via \\ Adaptive Feature-Wise Compression}
	\author{Yongjeong Oh, \IEEEmembership{Graduate Student Member,~IEEE}, Jaeho Lee, \IEEEmembership{Member,~IEEE}, \\ Christopher G. Brinton, \IEEEmembership{Senior Member,~IEEE}, and Yo-Seb Jeon, \IEEEmembership{Member,~IEEE}
	    \thanks{Yongjeong Oh, Jaeho Lee, and Yo-Seb Jeon are with the Department of Electrical Engineering, POSTECH, Pohang, Gyeongbuk 37673, Republic of Korea (e-mails: \{yongjeongoh, jaeho.lee, yoseb.jeon\}@postech.ac.kr).}
        \thanks{Christopher G. Brinton is with the Elmore Family School of Electrical and Computer Engineering, Purdue University, West Lafayette, IN 47907, USA (e-mail: cgb@purdue.edu).}
	}
	\vspace{-2mm}	
	
	\maketitle
	\vspace{-12mm}

	\begin{abstract} 
	    This paper proposes a novel communication-efficient split learning (SL) framework, named SplitFC, which reduces the communication overhead required for transmitting intermediate feature and gradient vectors during the SL training process. The key idea of SplitFC is to leverage different dispersion degrees exhibited in the columns of the matrices. SplitFC incorporates two compression strategies: (i) adaptive feature-wise dropout and (ii) adaptive feature-wise quantization. In the first strategy, the intermediate feature vectors are dropped with adaptive dropout probabilities determined based on the standard deviation of these vectors. 
        Then, by the chain rule, the intermediate gradient vectors associated with the dropped feature vectors are also dropped. In the second strategy, the non-dropped intermediate feature and gradient vectors are quantized using adaptive quantization levels determined based on the ranges of the vectors. To minimize the quantization error, the optimal quantization levels of this strategy are derived in a closed-form expression. Simulation results on the MNIST, CIFAR-100, and CelebA datasets demonstrate that SplitFC outperforms state-of-the-art SL frameworks by significantly reducing communication overheads while maintaining high accuracy. 
	\end{abstract}



	\section{Introduction}\label{Sec:Intro}

    Split learning (SL) is a privacy-enhancing distributed learning technique, in which multiple devices participate in the training of a global model (e.g., a neural network) through a parameter server (PS) \cite{SL,vepakomma2018split}. In a standard SL setup, the global model is divided into two sub-models: a device-side model, containing the first few layers of the global model stored on the device, and a server-side model, comprising the remaining layers stored on the PS. The SL training procedure operates in a round-robin fashion across devices. For every training iteration, each device performs forward propagation on its device-side model using a mini-batch stored exclusively on the device, generating an intermediate feature matrix of size (\textit{batch size}) $\times$ (\textit{the dimension of the intermediate features for a single input data})\footnote{In general, the intermediate map for one input data is represented as a 3-dimensional tensor of size (\textit{channel}) $\times$ (\textit{height}) $\times$ (\textit{width}). In this case, we can simply reshape the 3-dimensional tensor to a vector by concatenating the columns or rows of the tensor.}. This intermediate feature matrix is transmitted to the PS. Then the PS receives the intermediate feature matrix and continues forward propagation on the server-side model. After completing forward propagation for both models, the PS initiates backward propagation on the server-side model, producing an intermediate gradient matrix with the same size as the intermediate feature matrix. The device receives this intermediate gradient matrix and continues backward propagation on its device-side model. Upon finishing backward propagation for both models, gradients are calculated, and the global model is updated accordingly. This process is repeated for the next device.

    Federated learning (FL) is another representative distributed learning technique to enhance data privacy, in which the global model is trained by aggregating locally trained models on the devices \cite{Konecny:15,jeon2020compressive,sattler2019robust,R3_refer}. One of the key differences between SL and FL is the computational cost and storage requirement on devices \cite{thapa2022splitfed,li2020federated,han2022splitgp,letaief2021edge, tran2022privacy,CPSL}. In FL, all devices are responsible for training the entire global model, which can incur substantial computational and storage burdens especially when training large-scale models \cite{R3_refer,thapa2022splitfed,li2020federated,han2022splitgp}. In contrast, in SL, these burdens can be readily handled by adjusting the position of cut layer where the global model is divided into two sub-models \cite{letaief2021edge,tran2022privacy}. 
    Another key difference\footnote{In addition to these two distinctions, the privacy concerns associated with each learning approach also differ. For instance, in FL, data privacy issue may arise during the transmission of gradients and the updated global model. Conversely, in SL, this issue may emerge when exchanging intermediate matrices and device-side models.} is the information exchanged between the PS and the devices. In FL, each device transmits either the updated local model or the corresponding gradient, while the PS transmits the updated global model back to the devices. Conversely, in SL, each device transmits the intermediate feature matrix, and the PS transmits the intermediate gradient matrix. After this, each device transmits either the updated device-side model or the corresponding gradient, and the PS transmits the updated device-side model if necessary.

    A major bottleneck in SL is the substantial communication overhead$^2$ required for transmitting intermediate feature and gradient matrices during each iteration \cite{letaief2021edge, tran2022privacy}, which 
    is exacerbated as both the mini-batch size and the dimension of the intermediate features for a single input data increase.
    For example, if the wireless link capacity is 10 Mbps, and the global model has a batch size of 256 with an intermediate feature dimension of 8,192 for a single input, transmitting features and gradients over 100 iterations with 100 devices would take about $1.34 \times 10^5$ seconds. Such extensive communication times may not be suitable for low-power devices and practical SL applications, which often require operation at lower latencies.

    \subsection{Prior Works}
    To address the challenge of communication overhead in SL, various communication-efficient SL frameworks have been studied in the literature \cite{SL_ROUND_1,SL_ROUND_2,SL_NN_1,SL_SQ_1,RandTopK,SL_SQ_2}\footnote{It should be noted that while communication efficiency remains a crucial concern for SL, studies specifically addressing this issue are relatively scarce compared to the well-explored domain of FL.}. 
    The primary objective of these works is to compress intermediate features or gradients during the training procedure in SL. Three representative approaches to achieve this goal are (i) reducing the frequency of intermediate matrices transfer, (ii) incorporating autoencoders, and (iii) applying sparsification or quantization. 
    
    The first approach directly reduces the communication overhead in SL by decreasing the frequency of intermediate feature or gradient matrix transfer \cite{SL_ROUND_1,SL_ROUND_2}. In \cite{SL_ROUND_1}, the authors proposed a loss threshold that determines whether to exchange intermediate feature and gradient matrices at the PS. In \cite{SL_ROUND_2}, the authors introduced a local loss-based SL framework that avoids the need to send the intermediate gradient matrix from the PS to the devices, while making no changes in the transmission of the intermediate feature matrix from the devices to the PS. 
    The second approach employs a pre-trained autoencoder, where the encoder is inserted at the output of the device-side model, and the decoder is inserted at the input of the server-side model \cite{SL_NN_1}. This configuration enables the reduction of the number of columns for both intermediate feature and gradient matrices. 
    The third approach involves applying sparsification or quantization techniques to compress the intermediate feature and gradient matrices \cite{SL_SQ_1,RandTopK,SL_SQ_2}. In \cite{SL_SQ_1}, the authors leveraged the top-$S$ sparsification technique for both matrices. In \cite{RandTopK}, the authors enhanced the top-$S$ technique by introducing randomness to improve its performance. In \cite{SL_SQ_2}, the authors proposed a quantization technique in which the intermediate feature matrix is quantized based on ${\sf K}$-means clustering. 
    It is worth noting that the third approach is orthogonal to the previous two approaches; therefore, the sparsification or quantization techniques in the third approach can be combined with either of the first two approaches to further enhance communication efficiency.

    Although various communication-efficient SL frameworks \cite{SL_ROUND_1,SL_ROUND_2,SL_NN_1,SL_SQ_1,RandTopK,SL_SQ_2} have been developed, they still face significant challenges in enhancing communication efficiency without causing significant performance degradation. A primary limitation of existing methods is to apply uniform compression to all feature and gradient vectors across training iterations and devices. This can lead to less critical vectors being treated with the same priority as more informative ones, ultimately causing performance degradation, particularly at higher compression levels. Therefore, more flexible and adaptive compression techniques are needed to account for the varying importance of the features and gradients in SL, ensuring improved communication efficiency without compromising model performance.

    \subsection{Contributions}
    This paper presents a novel communication-efficient SL framework, named \textit{SplitFC}, which reduces the communication overhead of SL while mitigating any resulting 
    SL performance degradation. 
    The key idea of SplitFC is to adaptively compress intermediate feature and gradient vectors by leveraging dispersion levels that vary across the vectors. 
    Based on this idea, the presented framework incorporates two compression strategies: (i) adaptive feature-wise dropout and (ii) adaptive feature-wise quantization.    
    The major contributions of this paper can be summarized as follows:
    \begin{itemize}
        \item We propose the adaptive feature-wise dropout strategy, which probabilistically drops intermediate feature vectors with adaptive dropout probabilities.
        Moreover, our strategy also allows the PS to reduce the communication overhead for transmitting the intermediate gradient vectors: 
        by the chain rule, the PS only needs to transmit the intermediate gradient vectors associated with the non-dropped intermediate feature vectors.


        
        \item 
        We introduce an adaptive feature-wise quantization strategy for non-dropped intermediate feature and gradient vectors. Each vector is quantized using either a two-stage quantizer or a mean-value quantizer, depending on its range. By leveraging both quantizers, our strategy significantly reduces the communication overhead in SL.
        
        \item 
        To minimize the quantization error, we optimize the quantization levels allocated for both two-stage and mean value quantizers.
        This is achieved by 
        analytically characterizing the quantization errors, 
        and determining closed-form expressions for the optimal quantization levels based on the error analysis.
        
        \item Through extensive numerical evaluation of various image classification tasks in Sec.~VII, we demonstrate the superiority of SplitFC compared to state-of-the-art (SOTA) SL frameworks. Our results in Table~III of Sec.~VII also reveal that the combination of adaptive dropout and quantization outperforms each technique applied individually.
    \end{itemize}

    \section{Related Works}\label{Sec:Related_Works}
    In this section, we elaborate on the key distinctions between the proposed compression strategies and other compression methods explored in FL or centralized learning.

    \vspace{0.5mm}\noindent
    {\bf Sparsification and Dropout:} Sparsification methods aim to reduce the density of model parameters or gradients by transmitting only the non-zero or most significant elements \cite{Sparsification1,top_k1}. Dropout methods, commonly used for regularization, randomly drop individual features based on predefined dropout probabilities \cite{dp1,dp2}. In contrast, our proposed feature-wise dropout strategy is more adaptive, selectively discarding less informative feature vectors based on their standard deviation. This targeted approach ensures that only the most critical features are transmitted, significantly reducing communication overhead while maintaining model performance.
    

    \vspace{0.5mm}\noindent
    {\bf Scalar Quantization:} Scalar quantization, a widely used technique, involves quantizing each entry of model parameters, intermediate features, or gradient vectors individually  \cite{PowerQuant,EasyQuant,NoisyQuant,FedSQCS,SQTNNLS,Ternary,9773308,10216922,9837828,9399174}. 
    In contrast, our mean-value quantizer focuses on quantizing the mean of vectors rather than each individual entry. This method is particularly effective for small-range vectors, achieving a communication overhead of less than 1 bit per entry. Moreover, for vectors with larger ranges, we implement a two-stage quantizer that optimally allocates quantization bits based on the range of the vectors, thereby ensuring efficient bit usage and minimizing quantization error.
    
    \vspace{0.5mm}\noindent
    {\bf Vector Quantization:} Vector quantization involves grouping similar vectors into clusters and representing them with a shared codebook entry \cite{oh2022fedvqcs,UVeQFed}. Although our approach does not explicitly employ vector quantization, its concentration on the mean of the intermediate vector can be considered as a clustering technique. 
    Additionally, through the quantization of means, our strategy effectively improves communication efficiency in SL. 

    \vspace{0.5mm}\noindent
    {\bf Model Compression:}  Model compression, encompassing techniques such as knowledge distillation and pruning, targets a reduction in the neural network's size \cite{Knowledge_Distillation,Model_Prune,10227741,han2015deep,xu2024besa}. While these techniques focus on compressing model parameters, our approach in the SL paradigm targets a different challenge: reducing communication overhead by applying dropout and quantization to intermediate feature vectors. Additionally, although our dropout and quantization strategies do not directly reduce the network's size, combining them with model compression methods could provide new opportunities to enhance communication efficiency in SL. 
    
    \section{System Model}\label{Sec:System}
    We consider an SL system in which a PS and $K$ devices collaborate to train a global model (e.g., neural network) \cite{SL,vepakomma2018split}. In this system, the global model is divided into two sub-models: (i) the device-side model and (ii) the server-side model. The device-side model comprises the first few layers of the global model, while the server-side model consists of the remaining layers of the global model. The parameter vectors of these two sub-models are denoted by ${\bm w}_{\rm d} \in \mathbb{R}^{N_{\rm d}}$ and ${\bm w}_{\rm s} \in \mathbb{R}^{N_{\rm s}}$, respectively, where $N_{\rm d}$ and $N_{\rm s}$ represent the number of parameters for each model. The parameter vector of the entire global model is represented by ${\bm w} = [{\bm w}_{\rm d}^{\sf T}, {\bm w}_{\rm s}^{\sf T}]^{\sf T}$.
    
    In SL, it is common to assume that each device possesses a local training dataset $\mathcal{D}_k, \forall k \in \mathcal{K}=\{1,\ldots,K\}$, while the PS is not allowed to explicitly access local training datasets. Let $f({\bm w};{\bm u}_k)$ be a sample-wise loss function that quantifies how well the global model with parameter vector ${\bm w}$ fits a training data sample ${\bm u}_k\in\mathcal{D}_k,\forall k$. Given the nature of SL, where the global model is split into two sub-models, the sample-wise loss function $f$ can also be expressed as
    \begin{align}\label{eq:loss}
	    f({\bm w};{\bm u}_k) = h({\bm w}_{\rm s};g({\bm w}_{\rm d};{\bm u}_k)),
    \end{align}
    where $g$ is the device-side function that maps the input data to the feature space, and $h$ is the server-side function that maps the output of $g$ to a scalar loss value. Then, the local loss function for device $k$ is defined as $F_k({\bm w}) = \frac{1}{|\mathcal{D}_k|}\sum_{{\bm u}_k\in \mathcal{D}_k} f({\bm w};{\bm u}_k), \forall k$. Similarly, the global loss function is defined as
    \begin{align}\label{eq:global_loss}
        F({\bm w}) = \frac{1}{\sum_{j\in\mathcal{K}}|\mathcal{D}_j|}\sum_{k\in\mathcal{K}}|\mathcal{D}_k|F_k({\bm w}).
    \end{align}
    The primary goal of SL is to find the best parameter vector ${\bm w}^\star$ that minimizes the global loss function, i.e., ${\bm w}^\star = \argmin_{\bm w}F({\bm w})$.

    \subsection{A Typical SL Framework}\label{Sec:SL}
    In a standard SL system, to find the best parameter vector ${\bm w}^\star$, the global model is trained in a round-robin fashion across the devices, i.e., after one device participates in training, the next device engages. The training procedure between a device $k\in\mathcal{K}$ and the PS consists of forward and backward propagation processes, which are described below.
    \begin{itemize}
        \item {\bf Forward propagation:} Assume that device $k$ and the PS are engaged at iteration $t \in \{1,\ldots,T\}$, where $T$ is the total number of iterations. The parameter vectors of the device-side and server-side models are denoted by ${\bm w}_{\rm d}^{(t,k)}\in\mathbb{R}^{N_{\rm d}}$ and ${\bm w}_{\rm s}^{(t,k)}\in\mathbb{R}^{N_{\rm s}}$, respectively.
        A mini-batch for device $k$ at iteration $t$ is denoted by ${\bm U}^{(t,k)} = \{{\bm u}_1^{(t,k)},\ldots,{\bm u}_B^{(t,k)}\}$, where ${\bm u}_i^{(t,k)} \in \mathcal{D}_k$ and $B$ is the mini-batch size. 
        The \textit{forward propagation} process begins with the transmission of the parameter vector of the device-side model ${\bm w}_{\rm d}^{(t,k)}$. This transmission\footnote{In SL, in addition to transmitting intermediate matrices, the communication overhead for transmitting device-side model can be a crucial bottleneck. Fortunately, this problem can be effectively addressed using various model or gradient compression techniques \cite{UVeQFed,FedSQCS,Ternary,Sparsification1}. Note that these compression techniques do not consider or exploit the characteristics of intermediate matrices. This implies that a dedicated compression technique is still required to effectively compress the intermediate matrices.} can be achieved through  device-to-device communication with device $k-1$ \cite{D2D} or through relay-based communication involving device $k-1$, the PS, and device $k$ \cite{Relay}.
        Upon receiving this vector, device $k$ computes the intermediate feature matrix defined as
        \begin{align}\label{eq:features}
            {\bm F}^{(t,k)} = g({\bm w}_{\rm d}^{(t,k)};{\bm U}^{(t,k)}) \in \mathbb{R}^{B \times \bar{D}},
        \end{align}
        where $\bar{D}$ is the dimension of intermediate features for one input data. Each column of ${\bm F}^{(t,k)}$ is denoted by ${\bm f}_i^{(t,k)}, i\in\{1,\ldots,\bar{D}\}$ and each entry is denoted by $f_{b,i}^{(t,k)}, b\in\{1,\ldots,B\}$. After computing the intermediate feature matrix in \eqref{eq:features}, device $k$ sends it along with the corresponding labels to the PS. Upon receiving ${\bm F}^{(t,k)}$ and the corresponding labels, the PS continues the forward propagation and then computes the mini-batch loss defined as
        \begin{align}
            h({\bm w}_{\rm s}^{(t,k)};{\bm F}^{(t,k)}) &\triangleq \frac{1}{B}\sum_{i=1}^B f({\bm w}^{(t,k)};{\bm u}_i^{(t,k)}),
        \end{align}
        where ${\bm w}^{(t,k)} = \big[({\bm w}_{\rm d}^{(t,k)})^{\sf T}, ({\bm w}_{\rm s}^{(t,k)})^{\sf T}\big]^{\sf T}$.

    \begin{figure*}[hbt!]
        \centering \vspace{-3mm}{\epsfig{file=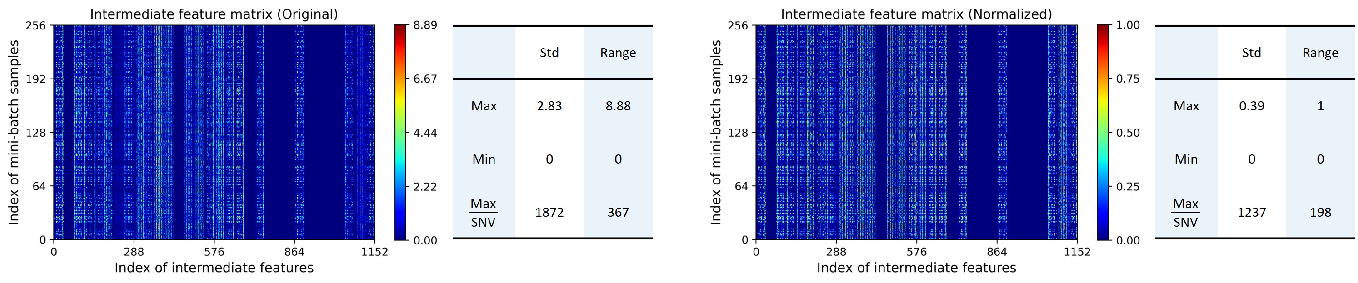,width=17cm}}\vspace{-2mm}
        \caption{Visualization of original and normalized intermediate feature matrices for the MNIST dataset with $B=256$ and $\bar{D}=1,152$. The terms Std and SNV refer to the standard deviation and the smallest non-zero values, respectively.}  \vspace{-3mm}
        \label{fig:Motivation}
    \end{figure*}
    	
        \item {\bf Backward propagation:} The \textit{backward propagation} process begins by computing a gradient vector with respect to the server-side model, which is represented by ${\bm g}_{\rm s}^{(t,k)} = \nabla_{{\bm w}_{\rm s}} h({\bm w}_{\rm s}^{(t,k)};{\bm F}^{(t,k)})$. The PS then computes the intermediate gradient matrix defined as
        \begin{align}\label{eq:gradient_features}
            {\bm G}^{(t,k)} = \nabla_{{\bm F}} h({\bm w}_{\rm s}^{(t,k)};{\bm F}^{(t,k)}) 
            \in \mathbb{R}^{B \times \bar{D}}.
        \end{align}
        After computing the intermediate gradient matrix in \eqref{eq:gradient_features}, the PS transmits it to device $k$. Upon receiving ${\bm G}^{(t,k)}$, device $k$ proceeds with the backward propagation based on the chain rule.
        As a result, device $k$ obtains a gradient vector with respect to the device-side model, namely ${\bm g}_{\rm d}^{(t,k)} = \nabla_{{\bm w}_{\rm d}} h({\bm w}_{\rm s}^{(t,k)};{\bm F}^{(t,k)})$. 
        Assuming that device $k$ and PS employ the SGD algorithm to update the parameter vectors of both device-side and server-side models, each parameter vector is updated as follows:
        \begin{align}\label{eq:update}
            {\bm w}_{p}^{(t,k+1)} &= {\bm w}_{p}^{(t,k)} - \eta{\bm g}_{p}^{(t,k)},~{p}\in\{{\rm d},{\rm s}\},
        \end{align}
        where $\eta$ is the learning rate.
        After updating the parameters, the parameter vector of the device-side model is transmitted to the next device, and then the forward propagation process for that device is initiated.
        If the index $k$ reaches $K$, the forward propagation process for device $1$ is initiated, while the index $(t, K+1)$ is replaced with $(t+1, 1)$.
        Meanwhile, when updating the device-side model using a momentum-based optimizer such as ADAM in \cite{ADAM}, the device $k$ can enhance the communication efficiency of SL by allowing the PS to update the device-side model without transmitting the updated model ${\bm w}_{\rm d}^{(t+1,k)}$ and momentum. This is possible because the PS can update the device-side model if it stores the first and second raw moments of the ADAM optimizer. 
    \end{itemize}
    \subsection{Key Challenge in SL} A key challenge in realizing the SL framework in Sec.~\ref{Sec:SL} is the significant communication overhead necessary for transmitting the intermediate feature and gradient matrices. While conventional compression techniques can help reduce this overhead, applying uniform compression to all vectors without considering their varying importance can lead to inefficiencies and performance degradation, especially at higher compression levels. Therefore, it is essential to develop more adaptive compression strategies that prioritize the most important features and gradients to improve communication efficiency and maintain model performance.

    \setlength{\textfloatsep}{7pt}
    \begin{algorithm}[t!]
        \setstretch{1.1}
        \caption{Communication-Efficient Split Learning via Adaptive Feature-Wise Compression (SplitFC)}\label{alg:Overall}
    	{\small
    	{\begin{algorithmic}[1]
            \REQUIRE ${\bm w}^{(1,1)}$, $R$, $C_{\rm e,d}$, $C_{\rm e,s}$
            \ENSURE ${\bm w}^{(T,K)}$
            \FOR {$t=1$ to $T$}
                \FOR {$k=1$ to $K$}
                    \vspace{-0.5mm}
                    \BULLET {\em At the device $k$:}
                        \STATE Receive ${\bm w}_{\rm d}^{(t,k)}$ from device $(k-1)$ or the PS
                        \STATE ${\bm F} = g({\bm w}_{\rm d}^{(t,k)};{\bm U}^{(t,k)})$
                        \STATE $\tilde{\bm F}, {\bm \delta} = {\sf FWDP}\big({\bm F}, R\big)$ (Algorithm~\ref{alg:FWDP})
                        \STATE ${\bm Q}_{\rm d}, {\bm \mu}_{\rm d} = {\sf FWQ}\big(\tilde{\bm F}, C_{\rm e,d}\big)$ (Algorithm~\ref{alg:FWQ})
                        \STATE Transmit ${\bm Q}_{\rm d}$, ${\bm \mu}_{\rm d}$, and ${\bm \delta}$ to the PS
                    \EndBULLET
                    \vspace{-0.7mm}
                    \BULLET {\em At the PS:}
                        \STATE Determine $\hat{\bm F}$ using ${\bm Q}_{\rm d}$, ${\bm \mu}_{\rm d}$, and ${\bm \delta}$ 
                        \STATE Compute the mini-batch loss $h({\bm w}_{\rm s}^{(t,k)}; \hat{\bm F})$
                        \STATE Compute the gradient vector ${\bm g}_{\rm s}$ for ${\bm w}_{\rm s}^{(t,k)}$
                        \STATE Update ${\bm w}_{\rm s}^{(t,k)}$ based on ${\bm g}_{\rm s}$
                        \STATE ${\bm G} = \nabla_{\hat{\bm F}} h({\bm w}_{\rm s}^{(t,k)}; \hat{\bm F})$
                        \STATE Determine $\tilde{\bm G}$ by dropping $\{{\bm g}_j\}_{\forall j\notin\mathcal{I}}$
                        \STATE ${\bm Q}_{\rm s}, {\bm \mu}_{\rm s} = {\sf FWQ}(\tilde{\bm G}, C_{\rm e,s})$ (Algorithm~\ref{alg:FWQ})
                        \STATE Transmit ${\bm Q}_{\rm s}$ and ${\bm \mu}_{\rm s}$ to device $k$
                    \EndBULLET
                    \vspace{-0.7mm}
                    \BULLET {\em At the device $k$:}
                        \STATE Determine $\hat{\bm G}$ using ${\bm Q}_{\rm s}$, ${\bm \mu}_{\rm s}$, and ${\bm \delta}$ 
                        \STATE Compute the gradient vector ${\bm g}_{\rm d}$ for ${\bm w}_{\rm d}^{(t,k)}$
                        \STATE Update ${\bm w}_{\rm d}^{(t,k)}$ based on ${\bm g}_{\rm d}$
                    \EndBULLET
                \ENDFOR
            \ENDFOR
    	\end{algorithmic}}}
    \end{algorithm}

    \section{Motivation of SplitFC}\label{Sec:Mot_Over}
    In this section, we present the motivation behind developing a novel communication-efficient SL framework, referred to as SplitFC. Our framework aims to effectively alleviate the communication overhead problem in SL by leveraging the characteristics of intermediate vectors.

    In a deep neural network, each intermediate layer can be thought of as a measurable feature, obtained by an output node of the device-side model from an attribute of the data sample \cite{lecun2015deep}. 
    As a result, some intermediate features may have similar values, indicating that the corresponding output nodes have captured common attributes in the mini-batch data samples. Conversely, some features may have significantly different values, suggesting that the corresponding output nodes have captured unique attributes. For example, consider SL for an image classification task using the MNIST dataset which consists of handwritten digit images. Certain output nodes of the device-side model may identify shared patterns such as loops or line segments common to multiple digits, leading to intermediate features with similar values. Other output nodes may distinguish between digits by recognizing unique attributes such as specific angles or intersections, resulting in intermediate features with dissimilar values.
    Consequently, intermediate features can exhibit varying levels of dispersion, which are typically measured by metrics such as standard deviation and range.

    \begin{figure*}[t!]
        \centering 
        {\epsfig{file=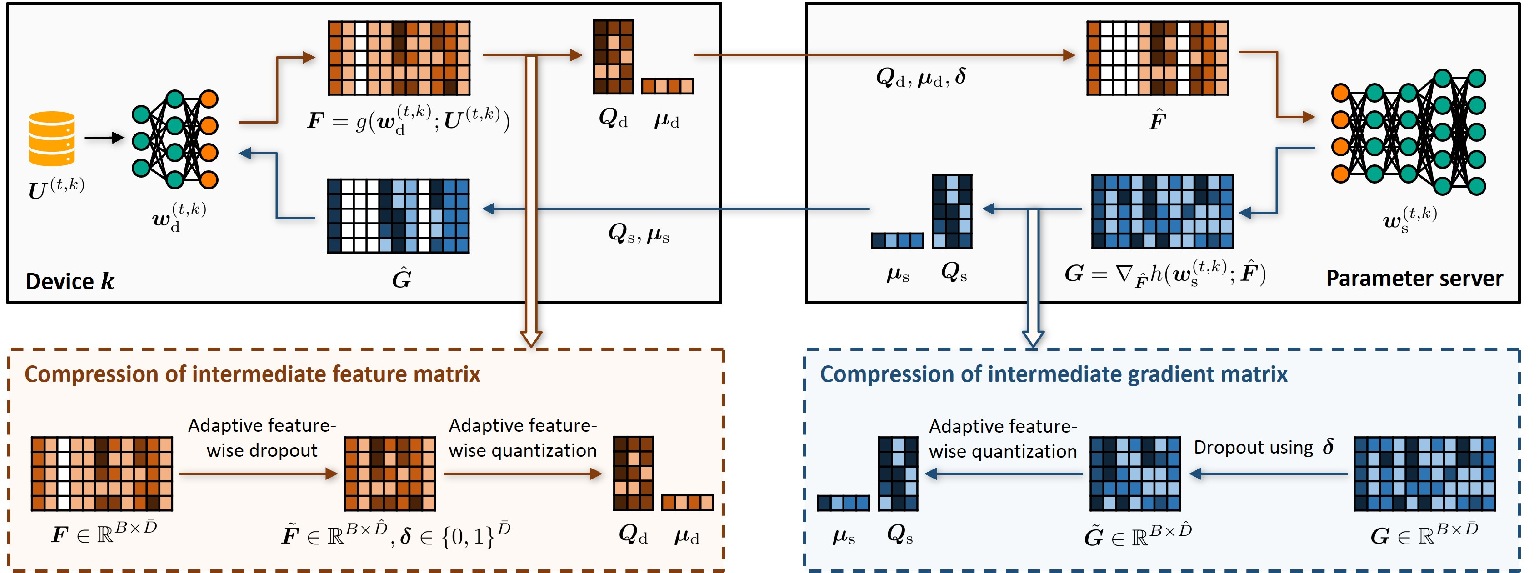,width=15.5cm}}\vspace{-1mm}
        \caption{Illustration of the proposed SL framework with adaptive feature-wise dropout and quantization strategies.}\vspace{-3mm}
        \label{fig:System}
    \end{figure*}
    
    To demonstrate the phenomenon described above, we present a simple numerical example in Fig.~\ref{fig:Motivation}. This example involves an image classification task using the MNIST dataset with $T=100$, and details of the simulation can be found in Sec.~\ref{sec:Sim}. In Fig.~\ref{fig:Motivation}, we highlight the minimum, maximum, and ratio between the maximum and smallest non-zero values for both the standard deviation and range. The left-hand side of Fig.~\ref{fig:Motivation} shows that the intermediate feature vectors exhibit significant differences in their values, standard deviations, and ranges. Fig.~\ref{fig:Motivation} also shows that some intermediate vectors have almost no changes in their values. These results align with our hypothesis on the dispersion levels in the intermediate feature vectors. In the right-hand side (RHS) of Fig.~\ref{fig:Motivation}, intermediate features are normalized to values between 0 and 1. Details of the normalization process will be elaborated in Sec.~\ref{sec:FWDP}. The RHS of Fig.~\ref{fig:Motivation} provides additional insight into the issue of feature dispersion, showing that even after normalization, intermediate feature vectors still exhibit different ranges and standard deviations. However, the overall disparity of their values, standard deviations, and ranges across different intermediate feature vectors is significantly reduced. 
    For instance, the original intermediate feature vectors with indices between $260$ and $288$ exhibit relatively low values. However, after normalization, these values become comparable to other large values.
    These results demonstrate the effectiveness of the normalization process in reducing the overall disparity of feature vectors and facilitating a fair comparison of their importance, as will be elaborated in Sec.~\ref{sec:FWDP}.

    Motivated by the discussion above, in SplitFC, we introduce adaptive feature-wise dropout and quantization strategies to compress the intermediate feature and gradient matrices. Unlike other methods that apply uniform compression, we leverage the varying dispersion levels in these vectors, enabling adaptive compression.
    The high-level procedure of SplitFC is illustrated in Fig.~\ref{fig:System} and summarized in {Algorithm~\ref{alg:Overall}}. Details on the adaptive feature-wise dropout and quantization strategies will be presented in Sec.~\ref{sec:FWDP} and Sec.~\ref{sec:FWQ}, respectively.

    \section{Adaptive Feature-Wise Dropout Strategy}\label{sec:FWDP}

    In this section, we present the adaptive feature-wise dropout strategy which aims at reducing the size of the intermediate feature and gradient matrices. The primary approach of this strategy is to probabilistically drop some of the intermediate feature vectors while retaining important ones with high probability. In the following, we provide the basic compression process of this strategy and then introduce the design of dropout probability for our strategy.


    \subsection{Basic Compression Process of Feature-Wise Dropout}
    In the proposed adaptive feature-wise dropout strategy, each intermediate feature vector ${\bm f}_i, i\in\{1,\ldots,\bar{D}\}$ is dropped probabilistically with a dropout probability $p_i$ during training. Any non-dropped intermediate feature vectors are then scaled by a factor of $\frac{1}{1-p_i}$ to ensure that the expected value of the scaled feature vector remains the same as that of the original feature vector. The resulting intermediate feature vector of this strategy is denoted by $\hat{\bm f}_i$ and can be expressed as 
    \begin{align}\label{eq:IF}
        \hat{\bm f}_i = \frac{\delta_i}{1-p_i}{\bm f}_i,
    \end{align}
    where $\delta_i\in\{0,1\}$ is a Bernoulli random variable with $\mathbb{P}(\delta_i = 1) = 1 - p_i$, indicating whether ${\bm f}_i$ is dropped or not.
    Let $\mathcal{I} = \{i: \delta_i = 1\}$ be the index set of non-dropped feature vectors
    and $\bar{\bm G} = \nabla_{\hat{\bm F}} h({\bm w}_{\rm s}; \hat{\bm F})\in\mathbb{R}^{B\times\bar{D}}$ be the intermediate gradient matrix computed at the PS. 
    It should be noted that when we drop an intermediate feature vector, the associated gradient vector naturally does not influence the subsequent backward propagation computations based on the chain rule. In this context, the intermediate gradient vector that needs to be transmitted to the device can be expressed as
    \begin{align}\label{eq:IG}
        \hat{\bm g}_j = 
        \begin{cases}
            \bar{\bm g}_j , & j \in \mathcal{I},\\
            {\bm 0}_{B}, & j \notin \mathcal{I}.
        \end{cases}
    \end{align}
    Let $\hat{D} = \sum_i \delta_i$ be the number of non-dropped feature vectors, $\tilde{\bm F}\in\mathbb{R}^{B\times\hat{D}}$ be the compressed intermediate feature matrix with each column being $\hat{\bm f}_j, j\in\mathcal{I}$, and $\tilde{\bm G}\in\mathbb{R}^{B\times\hat{D}}$ be the compressed intermediate gradient matrix with each column being $\hat{\bm g}_j, j\in\mathcal{I}$. 
    Then, the equations in \eqref{eq:IF} and \eqref{eq:IG} show that the device only needs to transmit the compressed intermediate feature matrix $\tilde{\bm F}$, while the PS only needs to transmit the compressed intermediate gradient matrix $\tilde{\bm G}$, provided that the device and PS share the information of the index vector ${\bm \delta}$\footnote{Because the PS can straightforwardly reconstruct the index set $\mathcal{I}$ using ${\bm \delta}$, it can determine $\hat{\bm F}$ from $\tilde{\bm F}$ and ${\bm \delta}$.}.

\begin{algorithm}[t]
    \setstretch{1.1}
    	\caption{Adaptive Feature-Wise Dropout Strategy}\label{alg:FWDP}
    	{\small
    	{\begin{algorithmic}[1]
    	    \PROC{${\sf FWDP} ({\bm F}, R)$}
                        \STATE $\displaystyle f_{\mathcal{I}_h}^{\rm max} = \max_{b\in\{1,\ldots, B\},i\in\mathcal{I}_h}f_{b,i},~ f_{\mathcal{I}_h}^{\rm min} = \min_{b\in\{1,\ldots, B\},i\in\mathcal{I}_h}f_{b,i},$
                        \Statex ~~~$\forall h\in\{1,\ldots,H\}$
                        \STATE ${f}_{b,j}^{\rm norm} = ({{f}_{b,j} - f_{\mathcal{I}_h}^{\rm min}}) / ({f_{\mathcal{I}_h}^{\rm max} - f_{\mathcal{I}_h}^{\rm min}}),$ 
                        \Statex ~~~$\forall b\in\{1,\ldots,B\},\forall j\in\mathcal{I}_h,\forall h$
                        \STATE $\mu_i = \frac{1}{B}\sum_{b=1}^B f_{b,i}^{\rm norm}, \forall i\in\{1,\ldots,\bar{D}\}$
                        \STATE $\sigma_i = \big\{\frac{1}{B}\sum_{b=1}^B ({f}_{b,i}^{\rm norm} - \mu_i)^2\big\}^{1/2}, \forall i$
                        \STATE $D = \bar{D}/R$
                        \STATE $q_i = ({\sigma_i D})/({\sum_{j=1}^{\bar{D}}\sigma_j}), \forall i$
                        \STATE $q^{\rm max} = \max_i q_i$
                        \vspace{1mm}
                        \STATE $p_i = 
                                \begin{cases}
                                    1 - q_i, & q^{\rm max} \leq 1 \\
                                    1 - \frac{(\sigma_i + C)D}{\sum_{j=1}^{\bar{D}}(\sigma_j + C)}, & q^{\rm max} > 1
                                \end{cases}, \forall i$ 
                        \vspace{1mm}
                        \STATE Generate the index vector ${\bm \delta}$ using $\{p_i\}_{\forall i}$
                        \STATE $\hat{\bm f}_i = {\delta_i}/({1-p_i}){\bm f}_i, \forall i$
                        \STATE $\mathcal{I} = \{i:\delta_i = 1\}$
                        \STATE Determine $\tilde{\bm F}$, with each column being $\hat{\bm f}_j,j\in\mathcal{I}$
                    \STATE \textbf{return} $\tilde{\bm F}$ and ${\bm \delta}$
                \ENDPROC
    	\end{algorithmic}}}
    \end{algorithm}

    \subsection{Design of Dropout Probability}\label{Sec:SS}
    To ensure that important features are preserved during the dropout process, we design the dropout probabilities by taking into account the standard deviations of the intermediate feature vectors.
    In our design, we prioritize intermediate feature vectors with high standard deviation to reduce the likelihood of dropping them.
    This prioritization is based on our assertion that intermediate feature vectors composed of dissimilar elements are likely to provide more valuable information for training the global model than those composed of similar elements. The rationale behind this assertion is that intermediate feature vectors composed of dissimilar elements contain unique attributes inherent in the training data and training the global model using these attributes can help to avoid overfitting to shared patterns in the training data, leading to better performance of the global model. In the following, we provide the details of our dropout probability design.
   
    It is worth nothing that comparing the importance of feature vectors based solely on raw standard deviation may not be fair, as the range of feature values can vary significantly across different feature vectors (see Fig.~\ref{fig:Motivation}(a)).
    To resolve this problem, we define a {\em normalized} feature as\footnote{In this normalization, when the intermediate layer is fully connected, the value of $H$ becomes equal to the number of nodes, which is denoted as $\bar{D}$. This approach is similar to batch normalization.}
    \begin{align}\label{eq:normalize}
        &\qquad\quad{f}_{b,j}^{\rm norm} = \frac{{f}_{b,j} - f_{\mathcal{I}_h}^{\rm min}}{f_{\mathcal{I}_h}^{\rm max} - f_{\mathcal{I}_h}^{\rm min}}, \nonumber \\ & j\in\mathcal{I}_h, b \in \{1,\ldots,B\}, h\in\{1,\ldots,H\},
    \end{align}
    where $H$ represents the number of channels of the original intermediate feature map, $\mathcal{I}_h$ is the index set of feature vectors corresponding to $h$-th channel 
    such that $\bigcup_{h=1}^H\mathcal{I}_h = \{1,\ldots,\bar{D}\}$ and $\mathcal{I}_{h_1}\cap\mathcal{I}_{h_2} = \emptyset,\forall h_1\neq h_2$. 
    The minimum and maximum values among the feature vectors in the $h$-th channel are denoted by $f_{\mathcal{I}_h}^{\rm min}$ and $f_{\mathcal{I}_h}^{\rm max}$, respectively i.e., $f_{\mathcal{I}_h}^{\rm min} = \min_{b\in\{1,\ldots, B\},i\in\mathcal{I}_h}f_{b,i}$ and $f_{\mathcal{I}_h}^{\rm max} = \max_{b\in\{1,\ldots, B\},i\in\mathcal{I}_h}f_{b,i}$. The normalization step in \eqref{eq:normalize} ensures that each feature vector has values between 0 and 1 (see Fig.~\ref{fig:Motivation}(b)), and allows a fair comparison of the importance of feature vectors.
    After the normalization, we compute the standard deviation of a normalized feature vector as follows:
    \begin{align}\label{eq:std}
        \sigma_i = \sqrt{\frac{1}{B}\sum_{b=1}^B \left({f}_{b,i}^{\rm norm} - \mu_i\right)^2},
    \end{align}
    where $i\in\{1,\ldots,\bar{D}\}$ and $\mu_i = \frac{1}{B}\sum_{b=1}^B f_{b,i}^{\rm norm}$.
    Let $D$ be the average number of non-dropped feature vectors. 
    Then, based on the standard deviation in \eqref{eq:std}, we define a constant $q_i$ as
    \begin{align}
        q_i = \frac{\sigma_i D}{\sum_{j=1}^{\bar{D}}\sigma_j},
    \end{align}
    which quantifies the relative importance of the feature vector ${\bm f}_i$ compared to the others.
    We finally determine the dropout probability $p_i$ for  the feature vector ${\bm f}_i$ as
    \begin{align}\label{eq:p_i_SS}
        p_i = 
        \begin{cases}
            1 - q_i, & q^{\rm max} \leq 1, \\
            1 - \frac{(\sigma_i + C_{\rm bias})D}{\sum_{j=1}^{\bar{D}}(\sigma_j + C_{\rm bias})}, & q^{\rm max} > 1,
        \end{cases}
    \end{align}
    where $q^{\rm max} = \max_{i\in\{1,\ldots,\bar{D}\}} q_i$, $\sigma^{\rm max} = \max_{i\in\{1,\ldots,\bar{D}\}} \sigma_i$ and $C_{\rm bias} \geq \frac{\sigma^{\rm max} D - \sum_j \sigma_j}{\bar{D}-D}$ is a hyper-parameter that serves as a bias to ensure that feature vectors with low standard deviation are selected while satisfying the probability axiom $0\leq p_i \leq 1$. The dropout probability in \eqref{eq:p_i_SS} shows that the larger the standard deviation, the smaller the dropout probability. This implies that intermediate feature vectors with high standard deviation are less likely to be dropped. 
    From this principle, our adaptive feature-wise dropout strategy prioritizes feature vectors with higher standard deviations, ultimately reducing the likelihood of dropping important feature vectors. We summarize the overall procedure of this strategy in {Algorithm~\ref{alg:FWDP}}. 


    \vspace{1mm}
    {\bf Remark 1 (Average communication overhead of adaptive feature-wise dropout strategy):}
    The expected value of $\hat{D}$ is determined as $\mathbb{E}[\hat{D}] = \mathbb{E}[\sum_i \delta_i] = \sum_i (1-p_i) = D$. Therefore, the average communication overhead required by the device for transmitting the intermediate feature matrix in \eqref{eq:IF} is determined as $C_{\rm d} = \frac{32B\bar{D}}{R} + \bar{D}$, where $R = \bar{D}/{D} > 1$ is a dimensionality reduction ratio that controls the degree of compression and the second term on the RHS represents the communication overhead for transmitting the index vector ${\bm \delta}$. Similarly, the average communication overhead required by the PS for transmitting the intermediate gradient matrix in \eqref{eq:IG} is determined as $C_{\rm s} = \frac{32B\bar{D}}{R}$. The communication overheads $C_{\rm d}$ and $C_{\rm s}$ show that the dimensionality reduction ratio $R$ controls both uplink and downlink communication overheads.
    

    \vspace{1mm}
    {\bf Remark 2 (Convergence analysis of the adaptive feature-wise dropout strategy):} In this analysis, we characterize the convergence rate of the adaptive feature-wise dropout strategy, following the approach in \cite{SL_SQ_2}. We begin by making the following assumptions for mini-batch SGD:

    {\em Assumption 1:} The objective function $F({\bm w})$ is $L$-Lipschitz smooth \big(i.e., $\|\nabla F({\bm w}_1) - \nabla F({\bm w}_2)\| \leq L \|{\bm w}_1 - {\bm w}_2\|, \forall {\bm w}_1, {\bm w}_2$\big).

    {\em Assumption 2:} The stochastic gradient has a bounded variance $\mathbb{E}[\|{\bm g}({\bm w}) - \nabla F({\bm w})\|^2] \leq \sigma^2$.

    Next, we determine the mean-squared error (MSE) of the adaptive feature-wise dropout strategy as
    \begin{align}\label{eq:MSE_FD}
        &\mathbb{E}\left[\big\|\hat{\bm F} - {\bm F}\big\|_{\rm F}^2\right] = \sum_{i=1}^{\bar{D}}\mathbb{E}[\|\hat{\bm f}_i - {\bm f}_i\|^2] \nonumber \\
        &\overset{(a)}{=} \sum_{i=1}^{\bar{D}}(\mathbb{E}[\|\hat{\bm f}_i\|^2] - \|{\bm f}_i\|^2) =\sum_{i=1}^{\bar{D}} \frac{p_i}{1-p_i}\|{\bm f}_i\|^2,
    \end{align}
    where the equality $(a)$ holds because $\mathbb{E}[\tilde{\bm f}_i] = {\bm f}_i,\forall i$.


    Then, based on the Assumptions 1 and 2 and the MSE in \eqref{eq:MSE_FD}, we can derive the convergence rate as follows:
    \begin{align}\label{eq:conv_rate}
        &\frac{1}{TK}\sum_{t=1}^T \sum_{k=1}^K \mathbb{E}\left[\left\|\nabla F({\bm w}^{(t,k)})\right\|^2\right] \nonumber \\
        &\leq \frac{4(F({\bm w}^{(1,1)}) - F({\bm w}^{\star}))}{\sqrt{TK}} + \frac{4L\sigma^2}{\sqrt{TK}} \nonumber\\&~~~~+ \gamma\left(\frac{4L}{\sqrt{TK}}+2\right) \max_{t,k}\sum_{i=1}^{\bar{D}} \frac{p_i^{(t,k)}}{1-p_i^{(t,k)}}\|{\bm f}_i^{(t,k)}\|^2,
    \end{align}
    where $\gamma = \max_{t,k}\big(\Lambda_1^{(t,k)}\big)^2+ \max_{t,k}\big(\Lambda_2^{(t,k)}\Lambda_3^{(t,k)}\big)^2$. Here, $\Lambda_1^{(t,k)}$, $\Lambda_2^{(t,k)}$, and $\Lambda_3^{(t,k)}$ are the largest eigenvalues of ${\partial^2 h({\bm w}_{\rm s}^{(t,k)} ; {\bm v}^{(t,k)} )}/{\partial {\bm w}_{\rm s}^{(t,k)}  \partial  {\bm v}^{(t,k)} }$, ${\partial^2 h({\bm w}_{\rm s}^{(t,k)} ;{\bm u}^{(t,k)} )}/{\partial ({\bm u}^{(t,k)})^2 }$, and ${\partial {\rm vec}({\bm F}^{(t,k)})}/{\partial {\bm w}_{\rm d}^{(t,k)}}$, respectively, where $ {\bm v}^{(t,k)}  = \lambda_1^{(t,k)}{\rm vec}({\bm F}^{(t,k)}) + (1-\lambda_1^{(t,k)}){\rm vec}(\hat{\bm F}^{(t,k)})$ for some $0<\lambda_1^{(t,k)}<1$ and $ {\bm u}^{(t,k)}  = \lambda_2^{(t,k)}{\rm vec}({\bm F}^{(t,k)}) + (1-\lambda_2^{(t,k)}){\rm vec}(\hat{\bm F}^{(t,k)})$ for some $0<\lambda_2^{(t,k)}<1$ \cite{SL_SQ_2}.


    The convergence rate in \eqref{eq:conv_rate} guarantees that the proposed framework, employing the adaptive feature-wise dropout strategy, converges to a neighborhood around the stationary point of the objective function. Moreover, the size of this neighborhood is in proportion to each dropout probability during training. Additionally, when the dropout probability approaches zero, this convergence rate aligns with that of mini-batch SGD.

    \section{Adaptive Feature-Wise Quantization Strategy}\label{sec:FWQ}
    The use of the adaptive feature-wise dropout strategy in Sec.~\ref{sec:FWDP} effectively reduces the size of the intermediate feature and gradient matrices that need to be transmitted by each device and the PS, respectively. This size reduction alone, however, may not be sufficient for achieving a significant reduction in the communication overhead required for transmitting intermediate features and gradients.
    To further reduce the communication overhead in SL, in this section, we present the adaptive feature-wise quantization strategy which reduces the number of quantization bits for representing the compressed intermediate feature and gradient matrices, $\tilde{\bm F}$ and $\tilde{\bm G}$.
    For the sake of simplicity, in the remainder of this section, we will use the term ``intermediate matrix'' denoted by ${\bm A}\in\mathbb{R}^{B\times \hat{D}}$, which is a generic term for the aforementioned matrices. We will also use the term ``intermediate vector'' denoted by ${\bm a}_i\in\mathbb{R}^{B}, i\in\{1,\ldots,\hat{D}\}$, representing $i$th column of ${\bm A}$.
    In addition, we shall assume that the intermediate vectors $\{{\bm a}_i\}_{i=1}^{\hat{D}}$ are sorted in descending order of the range without loss of generality.

\begin{algorithm}[t]
    \setstretch{1.1}
    	\caption{Adaptive Feature-Wise Quantization Strategy}\label{alg:FWQ}
    	{\small
    	{\begin{algorithmic}[1]
    	    \PROC{${\sf FWQ} ({\bm A}, C_{\rm ava})$}
                    \STATE $\displaystyle a_i^{\rm max} = \max_{b\in\{1,\ldots,B\}} a_{b,i}, a_i^{\rm min} = \min_{b\in\{1,\ldots,B\}} a_{b,i}, \forall i$
                    \STATE $\bar{a}_i = \frac{1}{B}\sum_{b=1}^Ba_{b,i}, \forall i$
                    \STATE Set $p^\prime$ to a large enough number
                    \FOR {$n=1$ to $N$}
                        \STATE $M = M_{N-n+1}$
                        \STATE $\displaystyle a^{\rm max} = \max_{j \in\{1,\ldots,M\}} a_j^{\rm max}, a^{\rm min} = \min_{j \in\{1,\ldots,M\}} a_j^{\rm min}$
                        \STATE $(\hat{a}_{u_j^{\rm min}},\hat{a}_{u_j^{\rm max}}) = {\sf Q}_{\rm ep}(a_j^{\rm min},a_j^{\rm max}), \forall j\in\{1,\ldots,M\}$
                        \STATE $\displaystyle \bar{a}^{\rm max} = \max_{k\in\{M+1,\ldots,\hat{D}\}} \bar{a}_k,~\bar{a}^{\rm min} = \min_{k\in\{M+1,\ldots,\hat{D}\}} \bar{a}_k$
                        \STATE Solve the problem $({\bf P})$ and determine $\{\hat{Q}_l\}_{l=0}^M$
                        \STATE $p = f(\hat{Q}_0,\ldots,\hat{Q}_M)$
                        \IF {$p>p^\prime$}
                            \STATE $\hat{Q}_l = \hat{Q}_l^\prime, \forall l$, and $M^\star = M^\prime$
                            \STATE {\bf break}
                        \ELSIF {$n=N$}
                            \STATE $M^\star = M$
                        \ELSE
                            \STATE $\hat{Q}_l^\prime = \hat{Q}_l, \forall l$, $M^\prime = M$, and $p^\prime = p$
                        \ENDIF
                    \ENDFOR
                    \STATE Quantize $\{{\bm a}_j\}_{j=1}^{M^\star}$ using the two-stage quantizer
                    \STATE Quantize $\{\bar{a}_k\}_{k={M^\star+1}}^{\hat{D}}$ using the mean-value quantizer
                    \STATE Determine ${\bm Q}$, with each column being ${\sf Q}_j({\bm a}_j)$
                    \STATE Determine ${\bm \mu}$, with each entry being ${\sf Q}_0(\bar{a}_k)$
                    \STATE {\bf return} ${\bm Q}$ and ${\bm \mu}$
                \ENDPROC
    	\end{algorithmic}}}
    \end{algorithm}

    \subsection{Quantizer Design}
    The fundamental idea of the adaptive feature-wise quantization strategy is to assign different quantization levels to intermediate vectors based on their ranges. Our key observation is that some intermediate vectors have almost no changes in their values, as can be seen in Fig.~\ref{fig:Motivation}; thereby, quantizing the means of these vectors, instead of their entries, can effectively reduce the communication overhead even without significant quantization error. Motivated by this observation, we employ two types of quantizers: (i) a two-stage quantizer, and (ii) a mean-value quantizer. The two-stage quantizer is used for quantizing intermediate vectors with the $M$ largest ranges, while the mean-value quantizer is used for quantizing the mean of each of the remaining $\hat{D}-M$ vectors. 

    

    \subsubsection{Two-stage quantizer}
    Our two-stage quantizer, used for quantizing the intermediate vectors with the $M$ largest ranges, works as follows: In the first stage, the maximum and minimum values of the intermediate vector are quantized using an {\em endpoint} quantizer. This process helps determine the upper and lower limits of the vector's entries using a lower bit budget.
    Then, in the second stage, the entries of the intermediate vector are quantized using an {\em entry} quantizer whose codebook is adpatively designed according the upper and lower limits determined by the endpoint quantizer.
    Let  ${\sf Q}_{\rm ep}: \mathbb{R}^2 \rightarrow \mathcal{C}_{\rm ep}^2$ be the endpoint quantizer which is common for the $M$ intermediate vectors, where $\mathcal{C}_{\rm ep}$ is the codebook of ${\sf Q}_{\rm ep}$ such that $|\mathcal{C}_{\rm ep}|=Q_{\rm ep}$, and $Q_{\rm ep}$ is the quantization level of ${\sf Q}_{\rm ep}$. 
    To determine the endpoint quantizer, the maximum and minimum values of the entries of the $M$ intermediate vectors are computed as
    \begin{align} 
        a^{\rm max} = \max_{j\in\{1,\ldots,M\}} a_{j}^{\rm max}~\text{and}~a^{\rm min} = \min_{j\in\{1,\ldots,M\}} a_{j}^{\rm min},
    \end{align}
    respectively, where $a_j^{\rm max} = \max_{b\in\{1,\ldots,B\}} a_{b,j}$ and $a_j^{\rm min} = \min_{b\in\{1,\ldots,B\}} a_{b,j}$. 
    Then, the $Q_{\rm ep}$-level uniform quantizer between $a^{\rm min}$ and $a^{\rm max}$ is employed as the endpoint quantizer, implying that $\mathcal{C}_{\rm ep} = \{\hat{a}_1,\ldots,\hat{a}_{Q_{\rm ep}}\}$ with 
    \begin{align}\label{eq:qu_end}
        \hat{a}_u = a^{\rm min} + (u-1)\Delta_{\rm ep}, u\in\{1,\ldots,Q_{\rm ep}\},
    \end{align}
    and $\Delta_{\rm ep} = \frac{a^{\rm max} - a^{\rm min}}{Q_{\rm ep} - 1}$.
    We assume that the quantization level $Q_{\rm ep}$ is shared by the PS and all the devices, while $a^{\rm min}$ and $a^{\rm max}$ are transmitted from the device to the PS. 
    Under this assumption, the same codebook $\mathcal{C}_{\rm ep}$ can be generated at both the PS and the devices  without explicitly exchanging the codebook itself. 
    Utilizing the above quantizer, the maximum and minimum values of the $j$-th intermediate vector ${\bm a}_j$ are quantized as ${\sf Q}_{\rm ep}(a_j^{\rm min},a_j^{\rm max}) = (\hat{a}_{u_j^{\rm min}},\hat{a}_{u_j^{\rm max}})$, where $u_j^{\rm min} = \left\lfloor {(a_j^{\rm min} - a^{\rm min})}/{\Delta_{\rm ep}} \right\rfloor +1$ and $u_j^{\rm max} = \left\lceil ({a_j^{\rm max} - a^{\rm min}})/{\Delta_{\rm ep}} \right\rceil + 1, j\in\{1,\ldots,M\}$. 
    By the definition, the outputs of the endpoint quantizer, $\hat{a}_{u_j^{\rm max}}$ and $\hat{a}_{u_j^{\rm min}}$, correspond to the upper and lower limits of the entries in ${\bm a}_j$, respectively, i.e., $\hat{a}_{u_j^{\rm min}} \leq a_{b,j} \leq \hat{a}_{u_j^{\rm max}}$, $\forall b\in\{1,\ldots,B\}$.
    This implies that a proper range for quantizing the entries in ${\bm a}_j$ can be specified by using $2 \log_2 Q_{\rm ep}$ bits.
    Thanks to this feature, our endpoint quantization with $\log_2 Q_{\rm ep}\ll 32$ requires a lower bit budget compared to explicitly transmitting the maximum and minimum values of ${\bm a}_j$. 
    In the second stage, the entry quantizer for the $j$-th intermediate vector is determined by utilizing its upper and lower limits ($\hat{a}_{u_j^{\rm max}}$ and $\hat{a}_{u_j^{\rm min}}$) determined by the endpoint quantizer. 
    Let ${\sf Q}_{j}: \mathbb{R}^B \rightarrow \mathcal{C}_{j}^B$ be the entry quantizer for the $j$-th intermediate vector  for $j \in\{1,\ldots,M\}$, where $\mathcal{C}_{j}$ is the codebook of ${\sf Q}_{j}$ such that $|\mathcal{C}_{j}|=Q_{j}$, and $Q_{j}$ is the quantization level of ${\sf Q}_{j}$.
    The $Q_{j}$-level uniform quantizer between $\hat{a}_{u_j^{\rm min}}$ and $\hat{a}_{u_j^{\rm max}}$ is employed as the entry quantizer for ${\bm a}_j$, implying that $\mathcal{C}_{j}$ consists of $Q_j$ values that are equally spaced between $\hat{a}_{u_j^{\rm min}}$ and $\hat{a}_{u_j^{\rm max}}$.
    Then, the entries of the $j$-th intermediate vector are quantized using the quantizer ${\sf Q}_j$ for $j \in\{1,\ldots,M\}$. 
    Let ${\bm Q}$ be a $B\times M$ matrix with each column being ${\sf Q}_j({\bm a}_j)$. Then, the bit budget for transmitting ${\bm Q}$ is given by $B\sum_{j=1}^M \log_2 Q_j$.

    \subsubsection{Mean-value quantizer}
    Our mean-value quantizer, used for quantizing the means of the intermediate vectors with the $\hat{D}-M$ smallest ranges, is designed using a uniform quantizer. 
    Let ${\sf Q}_0: \mathbb{R} \rightarrow {\mathcal{C}}_0$ be the mean-value quantizer which is common for the $\hat{D}-M$ intermediate vectors, where $\mathcal{C}_{0}$ is the codebook of ${\sf Q}_{0}$ such that $|\mathcal{C}_{0}|=Q_{0}$, and $Q_{0}$ is the quantization level of ${\sf Q}_{0}$. 
    To determine the mean-value quantizer, the mean of each of the $\hat{D}-M$ intermediate vectors, denoted by $\bar{a}_k = \frac{1}{B}\sum_{b=1}^Ba_{b,k}$, is computed for $k\in\{M+1,\ldots,\hat{D}\}$. 
    Utilizing the computed means, the maximum and minimum mean values are computed as $\bar{a}^{\rm max} = \max_{k\in\{M+1,\ldots,\hat{D}\}} \bar{a}_k$ and $\bar{a}^{\rm min} = \min_{k\in\{M+1,\ldots,\hat{D}\}} \bar{a}_k$, respectively.
    We assume that $\bar{a}^{\rm min}$ and $\bar{a}^{\rm max}$ are transmitted from the device to the PS. 
    Then, the $Q_0$-level uniform quantizer between $\bar{a}^{\rm min}$ and $\bar{a}^{\rm max}$ is employed as the mean-value quantizer, implying that ${\mathcal{C}}_0$ consists of $Q_0$ values that are equally spaced between $\bar{a}^{\rm min}$ and $\bar{a}^{\rm max}$.
    Utilizing this quantizer, the mean values in $\{\bar{a}_k\}_{k=M+1}^{\hat{D}}$ are quantized. 
    Let ${\bm \mu}$ be a $(\hat{D}-M)$-dimensional vector with each entry being ${\sf Q}_0(\bar{a}_k)$. Then, the bit budget for transmitting ${\bm \mu}$ is given by $(\hat{D}-M)\log_2 Q_0$.

    
    The total number of the quantization bits required by our adaptive feature-wise quantization strategy is given by 
    \begin{align}
        C_{\rm overhead} = &~2M\log_2Q_{\rm ep} + B\sum_{j=1}^M\log_2 Q_j    \nonumber \\
        &+ (\hat{D}-M)\log_2 Q_{0} + \hat{D} + 32\cdot 4. \label{eq:bit_overhead}
    \end{align}
    In \eqref{eq:bit_overhead}, the fourth term represents the bit budget for indicating whether each intermediate vector is quantized by the two-stage quantizer or not.
    The fifth term represents the bit budget for explicitly transmitting $a^{\rm max}$, $a^{\rm min}$, $\bar{a}^{\rm max}$, and $\bar{a}^{\rm min}$ using floating-point representation. 


    One prominent feature of our strategy is the use of the mean value quantizer, which quantizes the mean of the intermediate vector instead of quantizing its entries independently. 
    This quantizer can be an effective means of quantizing some intermediate vectors with almost no changes in their values, as can be seen from the numerical example in Fig.~\ref{fig:Motivation}. 
    Meanwhile, the mean value quantizer is {\em dimension-free} and therefore requires significantly fewer bits compared to an entry-wise quantizer.
    Another key feature of our strategy is the reduction of substantial bits required to determine a proper quantization range for each intermediate vector, by quantizing the maximum and minimum values of the vector using the endpoint quantizer.   
    The quantization error of our strategy can be minimized by optimizing the quantization levels and the number $M$ of the intermediate vectors that need to be quantized using the two-stage quantizer, which will be discussed in the following subsections. 
    The overall procedure of this strategy is summarized in {Algorithm~\ref{alg:FWQ}}.

    
    \vspace{1mm}
    {\bf Remark 3 (Relationship between feature-wise dropout and quantization):} The adaptive feature-wise dropout strategy reduces the size of the intermediate matrix by probabilistically dropping less informative feature vectors based on their standard deviations. However, even after dropout is applied, the remaining features must still be transmitted efficiently. This is where quantization plays a role, further compressing the remaining feature vectors based on their value range. The combination of these two techniques creates a synergistic effect: dropout reduces the number of features to be transmitted, while quantization further compresses the remaining feature vectors. Specifically, after dropout, some feature vectors with small ranges may still remain in the compressed intermediate feature matrix. These vectors can be effectively quantized via the mean-value quantizer. Meanwhile, even in cases where remaining feature vectors exhibit moderate or high ranges after the feature-wise dropout, the mean-value quantizer can still be effectively utilized. For instance, by quantizing feature vectors with moderate ranges using the mean-value quantizer, we can strategically allocate more bits to feature vectors with high ranges to minimize mean-squared error.

    \subsection{Quantization Level Allocation}\label{Sec:qla}
    To minimize the quantization error of the proposed adaptive feature-wise quantization strategy, we optimize the quantization levels of the two-stage quantizers and the mean value quantizer. 
    The quantization error of our strategy can be expressed as 
    \begin{align}\label{eq:qe}
        &\sum_{i=1}^{\hat{D}}\|{\bm a}_i - {\sf Q}_i({\bm a}_i)\|^2 = \sum_{i=1}^{{M}}{\sf QE}({\bm a}_i, Q_i) + \sum_{i=M+1}^{\hat{D}}{\sf QE}({\bm a}_i,\bar{a}_i),
    \end{align}
    where 
    ${\sf QE}({\bm a}_i, Q_i) = \|{\bm a}_i - {\sf Q}_i({\bm a}_i)\|^2$, and ${\sf QE}({\bm a}_i,\bar{a}_i) = \|{\bm a}_i - {\sf Q}_0(\bar{a}_i)\mathbbm{1}_B\|^2$. 
    Considering the quantization of endpoints in the two-stage quantizer, the first term on the RHS of \eqref{eq:qe} is upper-bounded as 
    \begin{align}\label{eq:SE_unif}
        {\sf QE}({\bm a}_j, Q_j) 
        \overset{(a)}{\leq} \frac{\tilde{a}_j^2 B}{4(Q_j-1)^2},
    \end{align}
    for all $j \in \{1,\ldots,M\}$, where $\tilde{a}_j = \hat{a}_{u_j^{\rm max}} - \hat{a}_{u_j^{\rm min}}$ and $(a)$ follows from the quantization error bound of uniform quantizer \cite{qerror}. 
    Considering the quantization of mean values in the mean-value quantizer, the second term on the RHS of \eqref{eq:qe} is upper-bounded as
    \begin{align}\label{eq:SE_mean}
        {\sf QE}({\bm a}_k,\bar{a}_k) &= \|{\bm a}_k - {\sf Q}_0(\bar{a}_k)\mathbbm{1}_B\|^2
        \nonumber \\
        &= \|{\bm a}_k - \bar{a}_k\mathbbm{1}_B + \bar{a}_k\mathbbm{1}_B - {\sf Q}_0(\bar{a}_k)\mathbbm{1}_B\|^2
        \nonumber \\
        &\leq 2\|{\bm a}_k - \bar{a}_k\mathbbm{1}_B\|^2 + 2\|\bar{a}_k\mathbbm{1}_B - {\sf Q}_0(\bar{a}_k)\mathbbm{1}_B\|^2
        \nonumber \\
        &\overset{(a)}{\leq} 2\|{\bm a}_k - \bar{a}_k\mathbbm{1}_B\|^2 + \frac{\tilde{a}_0^2B}{2(Q_0-1)^2},
        \nonumber \\
        &\overset{(b)}{\leq} \frac{(a_k^{\rm max} - a_k^{\rm min})^2 B}{2} + \frac{\tilde{a}_0^2B}{2(Q_0-1)^2},
    \end{align}
    for all $k \in \{M+1,\ldots,\hat{D}\}$, where $\tilde{a}_0 = \bar{a}^{\rm max}-\bar{a}^{\rm min}$, $(a)$ follows from the quantization error bound of uniform quantizer \cite{qerror} and $(b)$ follows from the result below.
    \begin{align}
        &\|{\bm a}_k - \bar{a}_k\mathbbm{1}_B\|^2 \nonumber \\
        &= \sum_{b=1}^B (a_{k,b} - a_k^{\rm min} + a_k^{\rm min} - \bar{a}_k)^2 \nonumber \\
        &= \sum_{b=1}^B (a_{k,b} - a_k^{\rm min})^2 - B (a_k^{\rm min} - \bar{a}_k)^2 \nonumber \\
        &\overset{(a)}{\leq} (a_k^{\rm max} - a_k^{\rm min})\sum_{b=1}^B (a_{k,b} - a_k^{\rm min}) - B (a_k^{\rm min} - \bar{a}_k)^2 \nonumber \\
        &= B(\bar{a}_k - a_k^{\rm min})\{(a_k^{\rm max} - a_k^{\rm min}) + (a_k^{\rm min} - \bar{a}_k)\} \nonumber \\
        &\overset{(b)}{\leq} \frac{(a_k^{\rm max} - a_k^{\rm min})^2B}{4},
    \end{align}
    where $(a)$ follows from $0 \leq (a_{k,b} - a_k^{\rm min}) \leq (a_{k}^{\rm max} - a_k^{\rm min})$, and $(b)$ follows from the arithmetic mean-geometric mean inequality. 
    
    Let $C_{\rm e,d}$ (bits/entry) and $C_{\rm e,s}$ (bits/entry) be the maximum uplink and downlink communication overheads allowed for transmitting each entry ${\bm F}$ and ${\bm G}$, respectively. Then, the total uplink and downlink communication overheads are determined as $B\bar{D}C_{\rm e,d}$ and $B\bar{D}C_{\rm e,s}$, respectively. Based on these overheads, the available bit budget $C_{\rm ava}$ for the adaptive feature-wise quantization strategy is determined as follows:
    \begin{enumerate}[(i)]
        \item If the device quantizes the compressed intermediate feature matrix $\tilde{\bm F}$, the available bit budget is given by $C_{\rm ava} = B\bar{D}C_{\rm e,d} - \bar{D}$, 
        where the second term on the RHS is the communication overhead for transmitting the index vector ${\bm \delta}$.
        \item If the PS quantizes the compressed intermediate gradient matrix $\tilde{\bm G}$, the available bit budget is given by $C_{\rm ava} = B\bar{D}C_{\rm e,s}$.
    \end{enumerate}
    
    We then formulate the quantization level allocation problem given the available communication overhead $C_{\rm ava}$, by substituting the results in \eqref{eq:SE_unif} and \eqref{eq:SE_mean} into \eqref{eq:qe} as follows:
    \begin{align}\label{eq:bit_allocation}
        ({\bf P})~&\min_{\{Q_i\}_{i=0}^M}~\bigg\{f(Q_0,\ldots,Q_M) = \sum_{i=1}^M \frac{\tilde{a}_i^2 B}{4(Q_i-1)^2}
        \nonumber \\
        &\,\,\,\, +\sum_{i=M+1}^{\hat{D}} \frac{(a_i^{\rm max} - a_i^{\rm min})^2 B}{2} 
        + \frac{\tilde{a}_0^2B(\hat{D}-M)}{2(Q_0-1)^2} \bigg\},
         \\
        &~\text{s.t.}~1 \leq \log_2 Q_l \leq 32, \forall l \in \{0,\ldots,M\},
        \label{eq:const_1} \\ 
        &\quad\,\,\,\,\,\,  C_{\rm overhead}\leq C_{\rm ava}, \label{eq:const_2}
    \end{align}
    where the constraint in \eqref{eq:const_1} specifies the maximum quantization level allowed for the entry quantizer for each intermediate vector. 
    The problem $({\bf P})$ is a convex optimization problem and belongs to the family of water-filling problems, specifically cave-filling problems \cite{cave_2}. Therefore, by utilizing the Karush-Kuhn-Tucker (KKT) conditions, we derive the optimal solution of the problem $({\bf P})$ as given in the following theorem:
    \begin{thm}\label{thm:Q_opt}
        The optimal solution of the problem $({\bf P})$ is
        \begin{align}\label{eq:Q_opt}
            Q_l^\star = \min\left\{2^{32}, \max\left\{2, \frac{(\frac{2}{3})^{\frac{1}{3}}u_l^\star}{v_l^\star} + \frac{v_l^\star}{2^{\frac{1}{3}}3^{\frac{2}{3}}} + 1 \right\} \right\},
        \end{align}
        for all $l\in\{0,\ldots,M\}$, where $u_0^\star = \frac{\tilde{a}_0^2B\log 2}{\nu^\star}$, $u_j^\star = \frac{\tilde{a}_j^2\log 2}{2\nu^\star}, j\in\{1,\ldots,M\}$, $v_l^\star = (u_l^\star \sqrt{81 - 12u_l^\star} + 9u_l^\star)^{\frac{1}{3}}, \forall l$, and $\nu^\star$ is the optimal Lagrange multiplier.
    \end{thm}
    \begin{IEEEproof}
        See Appendix~\ref{apdx:Q_opt}.
    \end{IEEEproof}
    \vspace{1mm}

    Theorem~\ref{thm:Q_opt} demonstrates that the quantizer for an intermediate vector with a large range is allocated with a higher quantization level than that for an intermediate vector with a small range.
    This implies that the quantization level allocation in \eqref{eq:Q_opt} naturally balances the quantization errors of intermediate vectors based on their ranges.
    Recall that the ranges of intermediate vectors can significantly differ as already discussed in Sec.~\ref{Sec:Mot_Over}.
    Therefore, our quantization strategy with the optimal quantization level allocation effectively reduces the overall quantization error by adaptively quantizing the intermediate vectors according to their ranges. 
    
    The optimal Lagrange multiplier $\nu^\star$ in Theorem~\ref{thm:Q_opt} can be easily obtained using various water-filling algorithms, such as the bisection search algorithm \cite{cave_2} and the fast water-filling algorithm \cite{FWF}. Once $\nu^\star$ is acquired through any water-filling algorithm, the optimal quantization level is determined as shown in \eqref{eq:Q_opt}. However, in practical scalar quantizers, the quantization level is restricted to an integer value that is greater than or equal to two, denoted by $\hat{Q}_l$. As a result, the optimal solution in \eqref{eq:Q_opt} might not be suitable for practical quantizers. One possible method to resolve this issue is to round down the solutions in \eqref{eq:Q_opt} to consistently satisfy the bit budget constraint in \eqref{eq:const_2} \cite{Du:20}. This approach has low computational complexity but may leave many unused bits. An alternative approach for better utilizing the available bit budget is to round off the solutions in \eqref{eq:Q_opt} and then adjust the quantization levels based on $C_{\rm ava}$ and the differences denoted by $d_j = B(\log_2 Q_j^\star - \log_2 \hat{Q}_j), \forall j$ and $d_0 = (\hat{D}-M)(\log_2 Q_0^\star - \log_2 \hat{Q}_0)$ \cite{round}. 
    This technique satisfies the bit budget constraint in \eqref{eq:const_2} while minimizing the number of unused bits. 
    
    The optimal quantization levels in Theorem~\ref{thm:Q_opt} can be determined not only at the device, but also at the PS.  
    This is because all the constants $\{\tilde{a}_l\}_{l=0}^M$ required for determining the optimal quantization levels can be computed at the PS using the information of $\bar{a}^{\rm max}$, $\bar{a}^{\rm min}$, and $\{\hat{a}_{u_j^{\rm min}}, \hat{a}_{u_j^{\rm max}}\}_{j=1}^M$ which are already assumed to be transmitted from the device. 
    Moreover, if the device explicitly transmits the optimal Lagrange multiplier $\nu^\star$, it is even not necessary to run the the water-filling algorithm at the PS.  
    It is worth noting that the communication overhead required for transmitting a single constant $\nu^\star$ is negligible compared to the total communication overhead $C_{\rm overhead}$. 
    These facts imply that both the device and the PS can generate the same optimal quantizers without explicitly exchanging the whole codebooks. 

    
    
    
    
    
    \subsection{Optimization of $M$}\label{Sec:oM}
    Utilizing the quantization error analysis and the optimal quantization levels presented in Sec.~\ref{Sec:qla}, 
    we also determine the best number $M^\star$ of the intermediate vectors that need to be quantized using the two-stage quantizer. 
    To achieve this, we consider a pre-defined candidate set for $M$, given by $\mathcal{M} = \{M_1,\ldots,M_N: 0\leq M_1 < \cdots < M_N \leq \hat{D}\}$.
    For each candidate $M\in\mathcal{M}$, we solve the problem $({\bf P})$, as described earlier, and evaluate the objective function in \eqref{eq:bit_allocation} using integer-valued quantization levels, i.e., $f(\hat{Q}_0,\ldots,\hat{Q}_M)$.
    We then determine the best number $M^\star$ by comparing the objective values and selecting the one that minimizes the objective function.
    To further reduce the computational complexity for determining $M^\star$, we can employ a stopping condition during the optimization process. For example, we can evaluate the objective values in descending order of $M$ and establish a stopping condition such that the evaluation halts if the current objective value is greater than the previous objective value. This method can save computation time, but still provides a reasonable estimate of $M^\star$. An outline of this method can be found in lines 12--21 of {Algorithm~\ref{alg:FWQ}}.

    By expanding the convergence analysis in {\bf Remark 2}, we can derive the convergence rate of the adaptive feature-wise quantization strategy while considering the quantization error in \eqref{eq:bit_allocation}. Although we omit the specific convergence rate for the sake of simplicity, it is worth noting that this convergence rate improves as the quantization error decreases. This result underscores the effectiveness of our quantization level allocation technique.
    

    \section{Simulation Results and Analysis}\label{sec:Sim}

    In this section, we demonstrate the superiority of SplitFC over existing SL frameworks, using simulations. The uplink and downlink compression ratios are defined as $32B\bar{D}/C_{\rm e,d}B\bar{D} = 32/C_{\rm e,d}$ and $32B\bar{D}/C_{\rm e,s}B\bar{D} = 32/C_{\rm e,s}$, respectively. Our simulations are conducted for the image classification task using three publicly accessible datasets: MNIST \cite{MNIST_v2}, CIFAR-100 \cite{CIFAR10}, and CelebA \cite{CelebA}. In our simulations, to consider various SL scenarios and provide a comprehensive analysis, we intentionally varied the number of devices $K$, the dimension of intermediate feature for one input data $\bar{D}$, and the batch-size $B$. Details of the learning scenarios for each dataset are described below. 
    \begin{itemize}
        \item {\em MNIST:}
        The global model is a simple variant of LeNet-5 in \cite{MNIST_v2}. The device-side model includes an input layer, a $3 \times 3$ convolutional layer with $16$ channels and one zero-padding, a $2 \times 2$ max pooling layer with a stride size of $2$, a $3 \times 3$ convolutional layer with $32$ channels, and a $2 \times 2$ max pooling layer with a stride size of $2$. The server-side model comprises a fully connected layer with $1,152$ nodes, followed by another fully connected layer with $128$ nodes, and finally an output layer with softmax activation. The activation function used for the hidden layers in both models is the rectified linear unit. The device-side model contains $N_{\rm d} = 4,800$ parameters, while the server-side model includes $N_{\rm s} = 148,874$ parameters. To update both models, the ADAM optimizer in \cite{ADAM} is adopted with an initial learning rate of $0.001$. 
        The data distribution is non-IID\footnote{In our simulations, the non-IID data distribution is considered, as it is a more realistic scenario in the context of SL. Additionally, due to this non-IID data distribution, SL may exhibit lower performance compared to centralized learning, where a global model is trained on the PS with access to the entire dataset \cite{non_IID}.}, where data samples with the same label are divided into subsets, and each device is assigned only two subsets, each containing different labels \cite{non_IID}.
        The numbers of devices and communication rounds are set to be $K=30$ and $T=200$, respectively. The dimension of intermediate features for one input data is determined as $\bar{D}=1,152$, and the mini-batch size is set to be $B=256$.
        
        \item {\em CIFAR-100:} 
        The global model is set to the ConvNeXt network pre-trained on the ImageNet dataset \cite{Liu_2022_CVPR}. 
        It has been divided at the $253$th layer so that the device-side and server-side models have a similar number of parameters, as described in \cite{SL_SQ_2,Sim_param}. To update both models, the ADAM optimizer in \cite{ADAM} with an initial learning rate $0.0001$ is adopted. The data distribution is non-IID, determined by a Dirichlet distribution with a concentration parameter $\beta = 0.3$ \cite{non_IID}. The numbers of devices and communication rounds are set to be $K=50$ and $T=100$, respectively. The dimension of intermediate features for one input data is determined as $\bar{D}=6,144$, and the mini-batch size is set to be $B=256$.
        
        
        \item {\em CelebA:} The global model is the MobileNetV3-Large network pre-trained on the ImageNet dataset \cite{howard2019searching}, which comprises 192 layers, and it is divided at the $82$nd layer. The device-side model contains $N_{\rm d} = 93,200$ parameters, while the server-side model includes $N_{\rm s} = 4,111,394$ parameters. To update both models, the ADAM optimizer in \cite{ADAM} with an initial learning rate $0.0001$ is adopted. Each local training dataset is determined by randomly grouping $20$ writers in CelebA dataset \cite{Model_Prune}. The numbers of devices and communication rounds are set to be $K=100$ and $T=40$, respectively. The dimension of intermediate features for one input data is determined as $\bar{D}=13,440$, and the mini-batch size is set to be $B=64$. 
        
    \end{itemize}
    For performance comparisons, we consider the following baselines:
    \begin{itemize}
        \item {\em Vanilla SL:} This framework assumes {\em lossless} transmission of the intermediate feature and gradient matrices with no compression, as described in Sec.~\ref{Sec:System}.
        
        \item {\em SplitFC:} This framework is the proposed SL framework summarized in Algorithm~\ref{alg:Overall}. In this framework, we set the hyper-parameter $C_{\rm bias}$ in \eqref{eq:p_i_SS} to $\frac{\sigma^{\rm max} D - \sum_j \sigma_j}{\bar{D}-D}$, the quantization level for endpoints to $Q_{\rm ep} = 200$, and the candidate set as $\mathcal{M} = \left\{D^{\rm max}{n}/{10}:n\in\{1,\ldots,10\}\right\}$, 
        where $D^{\rm max} = \min\big(\bar{D},\frac{C_{\rm ava} - 2\hat{D} - 32\cdot4}{B + 2\log_2 Q_{\rm ep} - 1}\big)$ represents the largest feasible value of $M$ for a specified bit budget $C_{\rm ava}$. Unless otherwise specified, we set the dimensionality reduction ratio $R$ to $16$. Meanwhile, we denote the SL framework using only the proposed adaptive feature-wise dropout as \textit{SplitFC-AD}.
        
        \item {\em SplitFC-Rand:} This framework is a simple modification of the adaptive feature-wise dropout strategy. In this framework, each intermediate feature vector ${\bm f}_i$ is randomly dropped with a dropout probability $p_i=1-{1}/{R},\forall i \in \{1,\ldots,\bar{D}\}$.
        
        \item {\em SplitFC-Deterministic:} This framework is an another modification of the adaptive feature-wise dropout strategy. In this framework, $(\bar{D}-D)$ intermediate feature vectors with small standard deviation in \eqref{eq:std} are dropped. 

        \item {\em FedLite:} This framework is the communication-efficient SL framework developed in \cite{SL_SQ_2}. In this framework, only the intermediate feature matrix is compressed using the ${\sf K}$-means algorithm, while the intermediate gradient matrix is transmitted with no compression. This compression technique can also be regarded as a type of vector quantization. The number of groups is set to one, and the number of subvectors is carefully selected as a constant that yields the highest classification accuracy among the divisors of $\bar{D}$.
        
        \item {\em RandTop-$S$:} This framework is the communication-efficient SL framework developed in \cite{RandTopK}. In this framework, each intermediate feature vector is sparsified by retaining only $S_{\rm d}$ entries. Corresponding intermediate gradient entries are dropped.
        The sparsification level $S_{\rm d}$ is determined as the largest integer $S_1$ such that $32S_1 + \log_2\big(\begin{smallmatrix} \bar{D}\\S_1 \end{smallmatrix} \big) \leq \bar{D}C_{\rm e,d}$. The hyper-parameter that controls the randomness is carefully selected from the set $\{0.1,0.2,0.3\}$. 
        
        \item {\em Top-$S$:} 
        This framework is an extension of the communication-efficient federated learning framework in \cite{SL_SQ_1}. 
        In this framework, the intermediate feature matrix is sparsified by retaining only the top-$S_{\rm d}$ entries with the largest magnitudes. Corresponding intermediate gradient entries are dropped. 
        
        \item {\em Scalar quantization:} We consider the SOTA scalar quantization techniques in \cite{PowerQuant,EasyQuant,NoisyQuant}, namely \textit{PowerQuant} (PQ), \textit{EasyQuant} (EQ), and \textit{NoisyQuant} (NQ) as baselines. Actually, the scalar quantization cannot achieve a communication overhead below 1 bit per entry without additional processing. To address this challenge, we incorporate the proposed adaptive feature-wise dropout strategy or the Top-$S$ sparsification strategy in \cite{SL_SQ_1} into the SOTA quantization techniques. This results in six different SL frameworks: (i) \textit{SplitFC-AD + PQ}, (ii) \textit{SplitFC-AD + EQ}, (iii) \textit{SplitFC-AD + NQ}, (iv) \textit{Top-$S$ + PQ}, (v) \textit{Top-$S$ + EQ}, and (vi) \textit{Top-$S$ + NQ}. Here, the sparsification level $S_{\rm d}$ for Top-$S$-based frameworks is determined as the largest integer $S_1\log_2\bar{Q} + \log_2\big(\begin{smallmatrix} \bar{D}\\S_1 \end{smallmatrix} \big) \leq \bar{D}C_{\rm e,d}$, where $\bar{Q}$ is set to an average quantization level of SplitFC-AD-based frameworks for fair comparison, that is $\bar{Q} = 2^{C_{\rm ava}R/B\bar{D}}$.
        
        
    \end{itemize}

    \begin{figure}[t]
        \centering\vspace{-1mm}
        {\epsfig{file=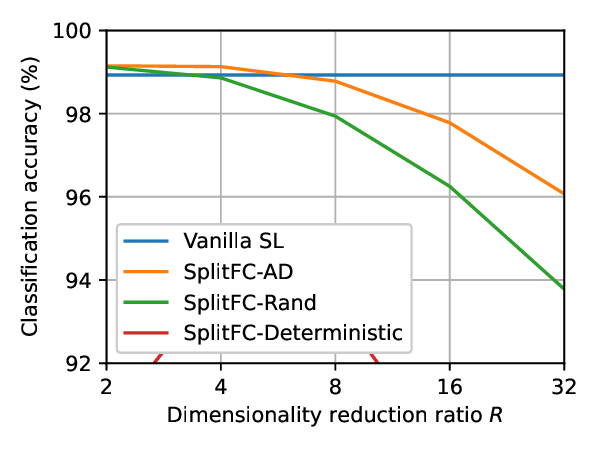,width=6.3cm}}\vspace{-4mm}
        \caption{Classification accuracy of SplitFC-based frameworks for the MNIST dataset with different values of $R$.}  
        \label{fig:RS_SS_v3}
    \end{figure}

    \begin{table*}[t]
        \renewcommand{\arraystretch}{1.0}
        \setlength{\tabcolsep}{3pt}
        \caption{Classification accuracy vs. uplink communication overhead for different SL frameworks.}\label{table:Acc_uplink}\vspace{-1mm}
        \footnotesize
        \centering
        \begin{tabular}{|c|c|c|c|c|c|c|c|c|c|c|c|c|} \hline
            Dataset & \multicolumn{4}{c|}{MNIST} & \multicolumn{4}{c|}{CIFAR-$100$} & \multicolumn{4}{c|}{CelebA} \\ \hline 
            \multicolumn{1}{|c|}{Uplink compression ratio} & $1\times$ & $160 \times$ & $240\times$ &$320\times$ & $1\times$ & $160 \times$ & $240\times$ &$320\times$ & $1\times$ & $160 \times$ & $240\times$ &$320\times$ \\ \hline 
            \multicolumn{1}{|c|}{$C_{\rm e,d}$ (bits/entry)} & $32$ & $0.2$ & $0.133$ &$0.1$ & $32$ & $0.2$ & $0.133$ &$0.1$ & $32$ & $0.2$ & $0.133$ &$0.1$ \\ \hline \hline
             Vanilla SL & $98.93$ & - & - & - & $73.20$ & - & - & - & $91.90$ & - & - & -  \\ \hline
            SplitFC & - & ${\bf 97.77}$ & ${\bf 97.53}$ & ${\bf 95.96}$ & - & ${\bf 71.16}$ & ${\bf 70.73}$ & ${\bf 67.69}$ & - & ${\bf 91.92}$ & ${\bf 91.92}$ & ${\bf 91.73}$ \\\hline
            FedLite & - & $85.25$ & $85.25$ & $70.19$ & - & $65.84$ & $61.08$ & $53.01$ & - & $75.93$ & $71.21$ & $64.78$ \\ \hline
            RandTop-$S$ & - & $84.90$ & $74.59$ & $69.91$ & - & $59.11$ & $56.60$ & $51.41$ & - & $55.75$ & $53.26$ & $52.60$ \\ \hline
            Top-$S$ & - & $79.05$ & $81.04$ & $78.28$ & - & $43.00$ & $39.17$ & $35.76$ & - & $53.60$ & $54.11$ & $52.48$ \\ \hline
            SplitFC-AD + PQ & - & $97.74$ & $96.60$ & $80.98$ & - & $71.09$ & $17.12$ & $11.80$ & - & $91.76$ & $59.19$ & $52.43$ \\ \hline
            SplitFC-AD + EQ & - & $97.77$ & $95.96$ & $82.35$ & - & $71.11$ & $66.00$ & $18.21$ & - & $91.91$ & $88.97$ & $88.49$ \\ \hline
            SplitFC-AD + NQ & - & $97.77$ & $95.90$ & $82.43$ & - & $71.11$ & $63.99$ & $12.41$ & - & $91.89$ & $88.95$ & $88.22$ \\ \hline
            Top-$S$ + PQ & - & $89.76$ & $87.26$ & $64.93$ & - & $62.17$ & $36.40$ & $35.83$ & - & $72.64$ & $51.74$ & $81.70$\\ \hline
            Top-$S$ + EQ & - & $90.60$ & $89.33$ & $80.11$ & - & $58.30$ & $59.13$ & $3.91$ & - & $68.77$ & $77.71$ & $62.02$ \\ \hline
            Top-$S$ + NQ & - & $90.36$ & $89.88$ & $81.60$ & - & $61.11$ & $60.58$ & $5.02$ & - & $67.68$ & $83.27$ & $60.37$ \\ \hline
        \end{tabular}
        \vspace{-3mm}
    \end{table*}
    
    In Fig.~\ref{fig:RS_SS_v3}, we compare the classification accuracies of SplitFC-based frameworks, namely SplitFC-AD, SplitFC-Rand, and SplitFC-Deterministic, for the MNIST dataset when the adaptive feature-wise quantization strategy is not applied. Fig.~\ref{fig:RS_SS_v3} demonstrates that SplitFC-AD exhibits greater robustness against the decrease in the dimensionality reduction ratio $R$ compared to SplitFC-Rand and SplitFC-Deterministic. This result shows the limitations of randomly and deterministically dropping feature vectors, which do not consider the importance and fairness of individual intermediate feature vectors during the training process, respectively. Furthermore, this result supports our assertion that intermediate feature vectors containing dissimilar elements provide more valuable information for training the global model than those composed of similar elements, as discussed in Sec.~\ref{Sec:SS}. Meanwhile, SplitFC-AD and SplitFC-Rand frameworks with low values of $R$ yield higher classification accuracy compared to the vanilla SL framework. This result suggests that the proposed adaptive feature-wise dropout strategy not only reduces the communication overhead but also has the potential to prevent neural networks from overfitting, as already reported in \cite{dp1,dp2}.

    \begin{table*}[t]
        \renewcommand{\arraystretch}{1.0}
        \setlength{\tabcolsep}{3pt}
        \caption{Classification accuracy vs. downlink communication overhead for different SL frameworks when the uplink communication overhead is $C_{\rm e,d} = C_{\rm e,s}/2$ bits/entry.}\label{table:Acc_downlink}\vspace{-2mm}
        \footnotesize
        \centering
        \begin{tabular}{|c|c|c|c|c|c|c|c|c|c|c|c|c|} \hline
            Dataset & \multicolumn{4}{c|}{MNIST} & \multicolumn{4}{c|}{CIFAR-$100$} & \multicolumn{4}{c|}{CelebA} \\ \hline 
            \multicolumn{1}{|c|}{Downlink compression ratio} & $1\times$ & $80 \times$ & $120\times$ &$160\times$ & $1\times$ & $80 \times$ & $120\times$ &$160\times$ & $1\times$ & $80 \times$ & $120\times$ &$160\times$ \\ \hline 
            \multicolumn{1}{|c|}{$C_{\rm e,s}$ (bits/entry)} & $32$ & $0.4$ & $0.266$ &$0.2$ & $32$ & $0.4$ & $0.266$ &$0.2$ & $32$ & $0.4$ & $0.266$ &$0.2$ \\ \hline \hline
             Vanilla SL ($C_{\rm e,d}=32$) & $98.93$ & - & - & - & $73.20$ & - & - & - & $91.90$ & - & - & -  \\ \hline
            SplitFC & - & ${\bf 97.77}$ & ${\bf 97.54}$ & ${\bf 95.98}$ & - & ${\bf 71.09}$ & ${\bf 70.56}$ & ${\bf 67.67}$ & - & ${\bf 91.97}$ & ${\bf 91.93}$ & ${\bf 91.76}$ \\\hline
            SplitFC-AD + PQ & - & ${\bf 97.77}$ & $96.53$ & $80.48$ & - & $71.04$ & $17.13$ & $12.13$ & - & $91.85$ & $57.98$ & $52.72$ \\ \hline
            SplitFC-AD + EQ & - & ${\bf 97.77}$ & $96.45$ & $82.51$ & - & $71.08$ & $65.48$ & $12.50$ & - & $91.93$ & $87.33$ & $77.98$ \\ \hline
            SplitFC-AD + NQ & - & ${97.76}$ & $96.50$ & $82.20$ & - & $71.06$ & $63.66$ & $14.82$ & - & $91.91$ & $88.01$ & $79.69$ \\ \hline
            Top-$S$ + PQ & - & $90.00$ & $87.51$ & $64.52$ & - & $61.06$ & $36.41$ & $35.75$ & - & $72.58$ & $51.32$ & $81.33$ \\ \hline
            Top-$S$ + EQ & - & $89.91$ & $89.76$ & $82.49$ & - & $61.26$ & $58.96$ & $3.35$ & - & $69.90$ & $77.73$ & $62.31$ \\ \hline
            Top-$S$ + NQ & - & $90.74$ & $89.27$ & $79.88$ & - & $58.34$ & $59.59$ & $4.06$ & - & $68.15$ & $82.86$ & $58.98$ \\ \hline
        \end{tabular}
        \vspace{-6mm}
    \end{table*}
    
    In Table~\ref{table:Acc_uplink}, we compare the classification accuracies of various SL frameworks for the MNIST, CIFAR-100, and CelebA datasets. In this simulation, we apply only uplink compression while assuming lossless transmission of the intermediate gradient matrix for all baselines. 
    Table~\ref{table:Acc_uplink} shows that the proposed SplitFC framework achieves the highest classification accuracy compared to other SL frameworks across all considered uplink communication overheads $C_{\rm e,d}$.
    In particular, the performance gap between SplitFC and the other baselines becomes more significant as the compression ratio increases. 
    Meanwhile, when combined with SplitFC-AD, scalar quantization techniques achieve significantly higher accuracy across most uplink communication overheads compared to when combined with Top-$S$. Furthermore, when combined with Top-$S$, the performance of scalar quantization techniques becomes unstable, especially for CelebA, where the accuracy decreases even as communication overhead increases. These results demonstrate the superiority of the proposed adaptive feature-wise dropout strategy and its robustness against reductions in uplink communication overhead.

    \begin{figure}[t]
        \centering
        {\epsfig{file=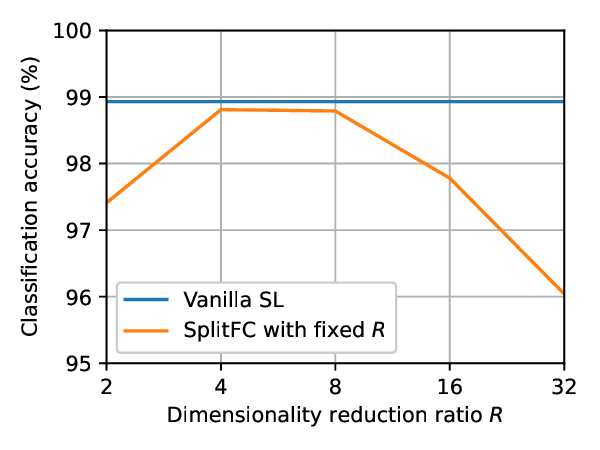,width=6.3cm}}\vspace{-5mm}
        \caption{Classification accuracy of SplitFC for the MNIST dataset with various choices of $R$ when $C_{\rm e,d} = 0.4$ bits/entry.}  \vspace{-1mm}
        \label{fig:R_opt_v3}
    \end{figure}

\begin{table*}[t]
    \renewcommand{\arraystretch}{1.1}
    \caption{Ablation Study on the Different Compression Strategies using the MNIST, CIFAR-100, and CelebA datasets.}\label{table:ablation}\vspace{-2mm}
    \footnotesize
    \centering
    \begin{tabular}{|m{1.5cm}|m{1.5cm}m{1.5cm}m{1.5cm}|wc{1.5cm}|wc{1.5cm}|wc{1.5cm}|wc{1cm}|} \hline
        & \multicolumn{1}{c}{Adaptive feature-} & \multicolumn{1}{c}{Two-stage} & \multicolumn{1}{c|}{Mean-value} & \multicolumn{3}{c|}{Classification accuracy (\%)} & \multicolumn{1}{c|}{Uplink and downlink} \\ \cline{5-7}
        & \multicolumn{1}{c}{wise dropout} & \multicolumn{1}{c}{quantizer} & \multicolumn{1}{c|}{quantizer} & \multicolumn{1}{c|}{MNIST} & \multicolumn{1}{c|}{CIFAR-100} & \multicolumn{1}{c|}{CelebA} & \multicolumn{1}{c|}{compression ratios} \\ \hline\hline 
        \multicolumn{1}{|c|}{Case 1}  & \multicolumn{1}{c}{\checkmark} &  &  & $93.46$ & $55.45$ & $91.23$ & $65\times$ \\ \hline
        \multicolumn{1}{|c|}{Case 2}  &  & \multicolumn{1}{c}{\checkmark} & \multicolumn{1}{c|}{\checkmark} & $90.37$ & $56.70$ & $88.98$ & $260\times$ \\ \hline
        \multicolumn{1}{|c|}{Case 3}  & \multicolumn{1}{c}{\checkmark} & \multicolumn{1}{c}{\checkmark} &  & $93.68$ & $46.49$ & $57.79$ & $260\times$ \\ \hline
        \multicolumn{1}{|c|}{Case 4}  & \multicolumn{1}{c}{\checkmark} & \multicolumn{1}{c}{\checkmark} & \multicolumn{1}{c|}{\checkmark} & $94.88$ & $66.28$ & $91.64$ & $260\times$ \\ \hline
    \end{tabular}\vspace{-3mm}
\end{table*}

    \begin{figure}[t]
        \centering
        {\epsfig{file=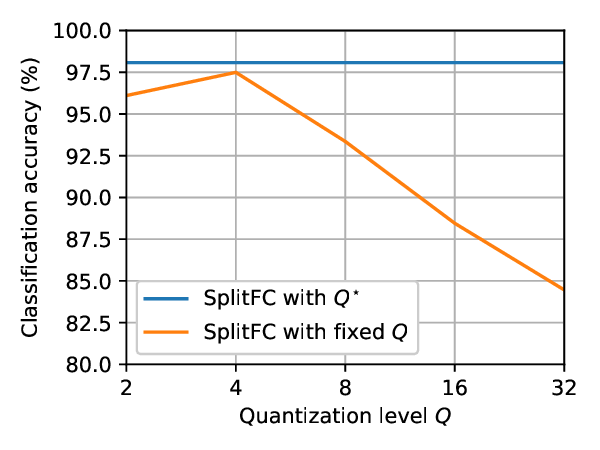,width=6.3cm}}\vspace{-4mm}
        \caption{Classification accuracy of SplitFC with and without the quantization level optimization with $C_{\rm e,d} = 0.2$ bits/entry and $R=8$.}  \vspace{-1mm}
        \label{fig:Opt_v2}
    \end{figure}
    
    In Table~\ref{table:Acc_downlink}, we evaluate the classification accuracy of various SL frameworks for MNIST, CIFAR-100, and CelebA datasets. In this simulation, we set the downlink compression ratios to 80, 120, and 160, and the uplink compression ratios to double those of the downlink compression ratios, considering that the transmission power of devices is generally lower than that of the PS \cite{Server_Edge}. Table~\ref{table:Acc_downlink} shows that the proposed SplitFC framework achieves the highest classification accuracy, consistently outperforming the other frameworks for all datasets.
    Moreover, Table~\ref{table:Acc_downlink} demonstrates that the classification accuracy of SplitFC does not degrade significantly even if the downlink communication overhead decreases. These results suggest that SplitFC effectively reduces the downlink communication overhead at the PS as well as the uplink communication overhead at the devices. Meanwhile, Tables~I and II show that the proposed dropout technique achieves higher performance when combined with the proposed quantization method compared to when it is combined with other existing quantization techniques. This highlights the effectiveness of combining the proposed dropout and quantization techniques.

    In Fig.~\ref{fig:R_opt_v3}, we evaluate the classification accuracy of SplitFC for the MNIST dataset by considering various choices of $R$. In this simulation, we assume lossless transmission of the intermediate gradient matrix and set the uplink communication overhead as $C_{\rm e,d} = 0.4$ bits/entry. Fig.~\ref{fig:R_opt_v3} shows that there exists an optimal selection of $R$ that yields the highest classification accuracy, even when the communication overhead remains constant (i.e., for a fixed $C_{\rm e,d}$). This phenomenon can be attributed to the fact that the performance of SplitFC relies on two types of errors: (i) dimensionality reduction error, and (ii) quantization error, both of which are determined by $R$. More specifically, as $R$ increases, the dimensionality reduction error also increases, resulting in performance degradation, as depicted in Fig.~\ref{fig:RS_SS_v3}. Conversely, when $R$ increases, the quantization error decreases because the dimension of the intermediate matrix needed for quantization becomes smaller, even though the available bit-budget stays constant. For example, the performance with $R=16$ is shown to be worse than that with $R=8$ because of the dominant dimensionality reduction error. Although we have observed that there exists an optimal selection of $R$ that balances the trade-off between dimensionality reduction error and quantization error, deriving an analytic expression for this optimal $R$ remains an open problem, which is an important area for future work.

    In Fig.~\ref{fig:Opt_v2}, we compare the classification accuracy of SplitFC with and without the quantization level optimization discussed in Sec.~\ref{Sec:qla} for the MNIST dataset. In this simulation, we assume lossless transmission of the intermediate gradient matrix and set the uplink communication overhead as $C_{\rm e,d} = 0.2$ bits/entry and $R=8$. Without quantization level optimization, we set $Q_l = Q \in \{2,4,8,16,32\}$ for all $l\in\{0,\ldots,D_Q^{\rm max}\}$. Here, $D_Q^{\rm max} = \min\big(\bar{D},\frac{C_{\rm ava} - \hat{D} - 32\cdot4 - \hat{D}\log_2 Q}{B\log_2 Q + 2\log_2 Q_{\rm ep} - \log_2 Q}\big)$ is the largest feasible value of $M$ for a specified bit budget $C_{\rm ava}$ and a fixed quantization level $Q$. Fig.~\ref{fig:Opt_v2} shows that SplitFC with the quantization level optimization achieves a $13.6\%$ improvement in classification accuracy compared to SplitFC with the worst-case parameter (i.e., $Q=32$ case). This result demonstrates the effectiveness of our quantization level optimization in enhancing the overall performance of SplitFC. 

    In Table~\ref{table:ablation}, we illustrate a comprehensive ablation study using the MNIST, CIFAR-100, and CelebA datasets to evaluate the impact of each proposed strategy in our framework. Table~\ref{table:ablation} shows that the combination of adaptive feature-wise dropout with the two-stage and mean-value quantizers (Case 4) achieves the highest classification accuracy across all datasets. In contrast, adaptive feature-wise dropout alone (Case 1), despite having a lower compression ratio, results in lower accuracy compared to Case 4. Similarly, the combination of only the two-stage and mean-value quantizers without dropout (Case 2) also falls short compared to Case 4. These results clearly demonstrate the limitations of applying either dropout or quantization individually, while emphasizing the necessity of combining both techniques to achieve high performance. Additionally, Table~\ref{table:ablation} shows that Case 4 outperforms Case 3. This highlights the effectiveness of the mean-value quantizer in optimizing bit usage, resulting in better compression and improved overall performance.
    

    \section{Conclusion}\label{Sec:Conclusion}
    In this paper, we have presented a communication-efficient SL framework that adaptively compresses intermediate feature and gradient matrices by exploiting various dispersion degrees exhibited in the columns of these matrices. In this framework, we have developed the adaptive feature-wise dropout and  quantization strategies, which significantly reduce both uplink and downlink communication overheads in SL. Furthermore, we have minimized the quantization error of the presented framework by optimizing quantization levels allocated to different intermediate feature/gradient vectors. An important direction for future research is to extend the presented framework to more complex communication scenarios, such as fading channels and device-specific heterogeneous conditions, to further enhance efficiency and scalability in real-world environments. Another promising direction is the joint optimization of cut layer positions and transmission strategies, alongside energy-efficient protocols, to maximize performance and resource utilization in SL under limited power resources. 



    \appendices
	\section{Proof of Theorem~\ref{thm:Q_opt}}\label{apdx:Q_opt}
	The Lagrangian function for \eqref{eq:bit_allocation} is given by
    \begin{align}\label{eq:lagrange}
        &\mathcal{L}(\{Q_i\}_{i=0}^M, \nu, \lambda_i, \gamma_i) 
        \nonumber \\
        &\!\!\!= \sum_{i=1}^M \frac{\tilde{a}_i^2 B}{4(Q_i-1)^2} +\sum_{i=M+1}^{\hat{D}} \frac{(a_i^{\rm max} - a_i^{\rm min})^2 B}{2} 
        \nonumber \\
        &\!\!\!+ \frac{\tilde{a}_0^2B(\hat{D}-M)}{2(Q_0-1)^2}
        +\sum_{i=0}^M\lambda_i (2-Q_i) + \sum_{i=0}^M\gamma_i(Q_i - 2^{32})
        \nonumber \\
        &\!\!\!+ \nu\bigg\{B\sum_{i=1}^M\log_2 Q_i + (\hat{D}-M)\log_2 Q_{0} + C_{\rm const} - C_{\rm ava} \bigg\},
    \end{align}
    where $C_{\rm const} = 2M\log_2 Q_{\rm ep} + \hat{D} + 32\cdot 4$. 
    Then, the KKT conditions for \eqref{eq:lagrange} are represented by
    \begin{gather}
        \label{KKT1}
         1 \leq \log_2 Q_i \leq 32,\forall i\in\{0,\ldots,M\}, \\
        \label{KKT2}
        \nu,\lambda_i,\gamma_i \geq 0, \forall i, \\
        \label{KKT3}
        \lambda_i (2-Q_i) = 0,\forall i, \\
        \label{KKT4}
        \gamma_i(Q_i-2^{32}) = 0, \forall i, \\
        \label{KKT5}
        \nu \bigg\{B\sum_{i=1}^M\log_2 Q_i +  (\hat{D}-M)\log_2 Q_{0} + C_{\rm const} - C_{\rm ava}\bigg\} = 0, \\
        \label{KKT6}
        \frac{\nu B}{Q_j\log 2} - \frac{\tilde{a}_j^2B}{2 (Q_j-1)^3} - \lambda_j + \gamma_j = 0, \forall j\in\{1,\ldots,M\}, \\
        \label{KKT7}
        \frac{\nu (\hat{D}-M)}{Q_0 \log 2} - \frac{\tilde{a}_0^2B(\hat{D}-M)}{(Q_{0}-1)^3} - \lambda_0 + \gamma_0 = 0.
    \end{gather}
    By multiplying \eqref{KKT6} with $(2-Q_j)$ and substituting \eqref{KKT3} into the resulting expression, we obtain the following equation:
    \begin{align}\label{two_condition1}
        \bigg(\frac{\nu B}{Q_j\log 2} - \frac{\tilde{a}_j^2B}{2 (Q_j-1)^3} + \gamma_j\bigg)(2-Q_j) = 0.
    \end{align}
    Under the condition in \eqref{KKT1}, two cases emerge from \eqref{two_condition1}:
    \begin{enumerate}[(i)]
        \item If ${Q_j} > 2$, then the equation $\frac{\nu B}{Q_j\log 2} - \frac{\tilde{a}_j^2B}{2 (Q_j-1)^3} + \gamma_j = 0$ holds. Then, using the condition $\gamma_j \geq 0$ from \eqref{KKT2}, we can deduce that $\frac{\tilde{a}_j^2B}{2 (Q_j-1)^3} - \frac{\nu B}{Q_j\log 2} \geq 0$. Solving this inequality with the consideration of $Q_j > 2$, we obtain
        \begin{align}\label{Q_j_cond1}
            \nu < \tilde{a}_j^2\log 2,~\text{for}~Q_j>2.
        \end{align}

        \item If $Q_j = 2$, then $\gamma_j = 0$ from \eqref{KKT4}. Substituting this into \eqref{KKT6} and using the condition $\lambda_j \geq 0$ from \eqref{KKT2}, we have
        \begin{align}\label{sol1}
            \nu \geq \tilde{a}_j^2\log 2,~\text{for}~Q_j=2.
        \end{align}
    \end{enumerate}
    Next, multiplying \eqref{KKT6} with $(Q_j - 2^{32})$ and substituting \eqref{KKT4} into the resulting expression, we have the following equation:
    \begin{align}\label{two_condition2}
        \bigg(\frac{\nu B}{Q_j\log 2} - \frac{\tilde{a}_j^2B}{2 (Q_j-1)^3} - \lambda_j\bigg)(Q_j - 2^{32}) = 0.
    \end{align}
    Under the condition in \eqref{KKT1}, two cases arise from \eqref{two_condition2}:
    \begin{enumerate}[(i)]
        \item If $Q_j < 2^{32}$, then the equation $\frac{\nu B}{Q_j\log 2} - \frac{\tilde{a}_j^2B}{2 (Q_j-1)^3} - \lambda_j = 0$ holds. Then, using the condition $\lambda_j \geq 0$ from \eqref{KKT2}, we can derive that $\frac{\nu B}{Q_j\log 2} - \frac{\tilde{a}_j^2B}{2 (Q_j-1)^3} \geq 0$. Solving this inequality with the consideration of $Q_j < 2^{32}$, we have 
        \begin{align}\label{Q_j_cond2}
            \nu > \frac{2^{31}\tilde{a}_j^2\log 2}{(2^{32}-1)^3},~\text{for}~Q_j<2^{32}.
        \end{align}

        \item If $Q_j = 2^{32}$, then $\lambda_j = 0$ from \eqref{KKT3}. Substituting this into \eqref{KKT6} and using the $\gamma_j \geq 0$ from \eqref{KKT2}, we have
        \begin{align}\label{sol2}
            \nu \leq \frac{2^{31}\tilde{a}_j^2\log 2}{(2^{32}-1)^3},~\text{for}~Q_j=2^{32}.
        \end{align}
    \end{enumerate}
    Next, consider the case $2 < Q_j < 2^{32}$. In this case, both $\lambda_j$ and $\gamma_j$ are equal to 0 according to conditions \eqref{KKT3} and \eqref{KKT4}, respectively. Substituting these values into the condition in \eqref{KKT6}, we obtain the following equation:
    \begin{align}\label{3rdeq}
        (Q_j-1)^3 = \frac{\tilde{a}_j^2\log 2}{2\nu} Q_j,~\text{for}~2<Q_j<2^{32}.
    \end{align}
    Given the results in \eqref{Q_j_cond1} and \eqref{Q_j_cond2}, we know that $\frac{1}{2} < \frac{\tilde{a}_j^2\log 2}{2\nu} < \frac{(2^{32}-1)^3}{2^{32}}$. Utilizing this inequality, we can derive the unique solution for \eqref{3rdeq} as follows:
    \begin{align}\label{sol3}
        Q_j = \frac{(\frac{2}{3})^{\frac{1}{3}}u_j}{v_j} + \frac{v_j}{2^{\frac{1}{3}}3^{\frac{2}{3}}} + 1,~\text{for}~2<Q_j<2^{32},
    \end{align}
    where $u_j = \frac{\tilde{a}_j^2\log 2}{2\nu}$ and $v_j = (u_j\sqrt{81 - 12u_j} + 9u_j)^{\frac{1}{3}}$.
    Finally, combining the results in \eqref{sol1}, \eqref{sol2}, and \eqref{sol3}, we obtain 
    \begin{align}
        Q_j = 
        \begin{cases}
          2, &\nu \geq \tilde{a}_j^2\log 2, \\
          \frac{(\frac{2}{3})^{\frac{1}{3}}u_j}{v_j} + \frac{v_j}{2^{\frac{1}{3}}3^{\frac{2}{3}}} + 1, &\frac{2^{31}\tilde{a}_j^2\log 2}{(2^{32}-1)^3} < \nu < \tilde{a}_j^2\log 2, \\
          2^{32}, &\nu \leq \frac{2^{31}\tilde{a}_j^2\log 2}{(2^{32}-1)^3}.
        \end{cases}
    \end{align}
    Similarly, we can determine the value of $Q_0$ using a procedure analogous to the one described above. Specifically, $Q_0$ is obtained as follows:
    \begin{align}
        Q_0 = 
        \begin{cases}
          2, &\nu \geq \tilde{a}_0^2 B \log 4, \\
          \frac{(\frac{2}{3})^{\frac{1}{3}}u_0}{v_0} + \frac{v_0}{2^{\frac{1}{3}}3^{\frac{2}{3}}} + 1, &\frac{2^{32}\tilde{a}_0^2 B \log 2}{(2^{32}-1)^3} < \nu < \tilde{a}_0^2 B \log 4, \\
          2^{32}, &\nu \leq \frac{2^{32}\tilde{a}_0^2\log 2}{(2^{32}-1)^3},
        \end{cases}
    \end{align}
    where $u_0 = \frac{\tilde{a}_0^2 B \log 2}{\nu}$ and $v_0 = (u_0\sqrt{81 - 12u_0} + 9u_0)^{\frac{1}{3}}$. The Lagrange multiplier $\nu$ must fulfill the condition in \eqref{KKT5}, and the optimal Lagrange multiplier $\nu^\star$ can be obtained using a variety of water-filling algorithms found in the literature \cite{cave_2, FWF}. Once $\nu^\star$ is obtained, the optimal quantization level can be determined as shown in \eqref{eq:Q_opt} of Theorem~\ref{thm:Q_opt}.


    \bibliographystyle{IEEEtran}
    \bibliography{Reference}
    
\end{document}